\documentclass[iop]{emulateapj}
                                                                                                                     
\usepackage{epsfig}
\usepackage{natbib}
\usepackage{graphicx}
\usepackage{amsmath}
\usepackage{multirow}
\usepackage{float}
\def\simgt{\lower.5ex\hbox{$\; \buildrel > \over \sim \;$}}
\def\simlt{\lower.5ex\hbox{$\; \buildrel < \over \sim \;$}}

\def\amin{\ifmmode^{\prime}\else$^{\prime}$\fi}
\def\asec{\ifmmode^{\prime\prime}\else$^{\prime\prime}$\fi}

\def\simgt{\lower.5ex\hbox{$\; \buildrel > \over \sim \;$}}
\def\simlt{\lower.5ex\hbox{$\; \buildrel < \over \sim \;$}}

\newcommand\chandra{{\it Chandra}}

\newcommand\xmm{{\it XMM-Newton}}
\newcommand\XMM{{\it XMM-Newton}}
\newcommand\integral{{\it INTEGRAL}/IBIS}

\newcommand\nustar{{\it NuSTAR\/}}

\newcommand\suzaku{{\it Suzaku\/}}

\def\sgrb{Sgr~B2}

\shorttitle{NuSTAR Observations of Sgr~B2}
\shortauthors{S. Zhang et al.}

\begin{document}

\title{Hard X-ray Morphological and Spectral Studies of The Galactic Center Molecular Cloud Sgr B2: Constraining Past Sgr~$\rm A^\star$ Flaring Activity}

\author{Shuo Zhang\altaffilmark{1}, Charles J. Hailey\altaffilmark{1}, Kaya Mori\altaffilmark{1}, Ma\"{i}ca Clavel\altaffilmark{2}, R\'{e}gis Terrier\altaffilmark{3}, Gabriele Ponti\altaffilmark{4}, \\
Andrea Goldwurm\altaffilmark{2,3}, Franz E. Bauer\altaffilmark{5,6,7}, Steven E. Boggs\altaffilmark{8},  William W. Craig\altaffilmark{8,9}, \\ Finn E. Christensen\altaffilmark{10}, Fiona A. Harrison\altaffilmark{11}, Jaesub Hong\altaffilmark{12},
Melania Nynka\altaffilmark{1}, Daniel Stern\altaffilmark{13}, \\ Simona Soldi\altaffilmark{3}, John A. Tomsick\altaffilmark{8} and William W. Zhang\altaffilmark{14}}

\altaffiltext{1}{Columbia Astrophysics Laboratory, Columbia University, New York, NY 10027, USA; shuo@astro.columbia.edu}
\altaffiltext{2}{Service d'Astrophysique/IRFU/DSM, CEA Saclay, 91191 Gif-sur-Yvette Cedex, France}
\altaffiltext{3}{Unit\'{e} mixte de recherche Astroparticule et Cosmologie, 10 rue Alice Domon et L\'{e}onie Duquet, 75205 Paris, France} 
\altaffiltext{4}{Max-Planck-Institut f\"{u}r extraterrestrische Physik, Giessenbachstrasse 1, D-85748, Garching bei M\"{u}nchen, Germany}
\altaffiltext{5}{Instituto de Astrof\'{\i}sica, Facultad de F\'{i}sica, Pontificia Universidad Cat\'{o}lica de Chile, 306, Santiago 22, Chile}
\altaffiltext{6}{Millennium Institute of Astrophysics, Chile} 
\altaffiltext{7}{Space Science Institute, 4750 Walnut Street, Suite 205, Boulder, CO 80301, USA}
\altaffiltext{8}{Space Sciences Laboratory, University of California, Berkeley, CA 94720, USA}
\altaffiltext{9}{Lawrence Livermore National Laboratory, Livermore, CA 94550, USA}
\altaffiltext{10}{DTU Space - National Space Institute, Technical University of Denmark, Elektrovej 327, 2800 Lyngby, Denmark}
\altaffiltext{11}{Cahill Center for Astronomy and Astrophysics, California Institute of Technology, Pasadena, CA 91125, USA}
\altaffiltext{12}{Harvard-Smithsonian Center for Astrophysics, Cambridge, MA 02138, USA}
\altaffiltext{13}{Jet Propulsion Laboratory, California Institute of Technology, Pasadena, CA 91109, USA}
\altaffiltext{14}{X-ray Astrophysics Laboratory, NASA Goddard Space Flight Center, Greenbelt, MD 20771, USA}

\begin{abstract}

Galactic Center (GC) molecular cloud \sgrb\ is the best manifestation of an X-ray reflection nebula (XRN) reprocessing a past giant outburst from the supermassive black hole Sgr~A$^\star$.
Alternatively, Sgr~B2 could be illuminated by low-energy cosmic ray electrons (LECRe) or protons (LECRp). 
In 2013, \nustar\ for the first time resolved Sgr~B2 hard X-ray emission on sub-arcminute scales.
Two prominent features are detected above 10~keV -- a newly emerging cloud G0.66$-$0.13 and the central 90\asec\ radius region containing two compact cores Sgr~B2(M) and Sgr~B2(N) surrounded by diffuse emission.
It is inconclusive whether the remaining level of Sgr~B2 emission is still decreasing or has reached a constant background level.
A decreasing Fe K$\alpha$ emission can be best explained by XRN while a constant background emission can be best explained by LECRp.
In the XRN scenario, the 3--79~keV Sgr~B2 spectrum can well constrain the past Sgr~A$^{\star}$ outburst, resulting in an outburst spectrum with a peak luminosity of  $L_{3-79\rm~keV} \sim 5\times10^{38} \rm~erg~s^{-1}$ derived from the maximum Compton-scattered continuum and the Fe K$\alpha$ emission consistently. 
The XRN scenario is preferred by the fast variability of G0.66$-$0.13, which could be a molecular clump located in the \sgrb\ envelope reflecting the same Sgr~A$^{\star}$ outburst. 
In the LECRp scenario, we derived the required CR ion power $dW/dt=(1-4)\times10^{39}\rm~erg~s^{-1}$ and the CR ionization rate $\zeta_{H}=(6-10)\times 10^{-15}\rm~H^{-1}~s^{-1}$.
The Sgr~B2 background level X-ray emission will be a powerful tool to constrain GC CR population.

\end{abstract}

\section{Introduction}

The Galactic center (GC) supermassive black hole Sagittarius~A$^\star$ (Sgr~A$^\star$) is the closest such object and is thus an ideal target for investigation of galactic nuclei and their activity cycles \citep{Morris1999, Ponti2013}.
Sgr~A$^\star$ is an underluminous black hole with a bolometric luminosity about $10^{-9}$ times the Eddington luminosity for a $4 \times 10^{6} M_\sun$ black hole \citep{Ghez2008, Gillessen2009}.
Its current X-ray quiescent state, with a luminosity of $L_{X} \sim 10^{33} \rm~erg~s^{-1}$ \citep{Baganoff2003}, is punctuated by flares up to a few times $10^{35}~\rm erg~s^{-1}$ (e.g. \citealp{Baganoff2001,Porquet2008, Nowak2012}), during which the bolometric luminosity is still orders of magnitudes lower than its Eddington luminosity.
Hard X-rays up to $\sim79$~keV have also been detected from the flares by \nustar~\citep{Barriere2014}.
Whether it has ever experienced more substantial increases in activity as observed in low-luminosity Active Galactic Nuclei (AGN) is still under discussion.

Indication of such past activity of Sgr~A$^\star$ has come from Galactic center molecular clouds (GCMCs).
In the populous Central Molecular Zone (CMZ, \citealt{Morris1996}), \sgrb\ is the densest and most massive molecular cloud.  
It has complicated substructures including compact star-forming cores like Sgr~B2(M), Sgr~B2(N) and Sgr~B2(S) (\citealp{Etxaluze2013} and references therein).
In an extremely simplified picture, its density distribution can be described as a dense core with a radius of 0.15--0.3~pc (2\asec--4\asec\ assuming the Sgr B2 distance of $7.9\pm0.8$~kpc, \citealp{Reid2009}) and an $\rm H_{2}$ density of (3--9)$\times 10^{6} \rm ~cm^{-3}$, surrounded by an envelope extended to $\sim 5$~pc ($\sim 2.2\amin$) with a density of $10^{4}$--$10^{5} \rm ~cm^{-3}$ and a larger diffuse component reaching $\sim22.5$~pc with a roughly constant density of $\sim 10^{3} \rm ~cm^{-3}$ \citep{Lis1990, DeVicente1997}.

\sgrb\ is the first GCMC from which a strong 6.4~keV Fe K$\alpha$ line was discovered \citep{Koyama1996}. 
{\it ASCA} detected this significant line feature with an equivalent width of about 1~keV, which was later confirmed by \chandra, \suzaku\ and \xmm\ observations \citep{Murakami2001, Koyama2007, Terrier2010}.
Time variability of the \sgrb\ Fe K$\alpha$ line was revealed by years of monitoring by different instruments.
The 6.4~keV line flux began declining in 2001 and decreased by a factor of $\sim0.4$ by 2005 \citep{Inui2009}, and further decreased by a factor of $0.39\pm0.04$ from 2005 to 2009 \citep{Nobukawa2011}.
The flux and morphology change of the X-ray emission is also observed from other GC molecular clouds.
Interestingly, the clouds exhibit Fe K$\alpha$ line variability in different ways, some rising, others are decreasing and an emission peak is detected in the ``Bridge" structure with a time scale as short as two years \citep{Muno2007, Ponti2010, Capelli2012, Clavel2013}. 

Reflection of incoming X-rays by cold molecular material is a natural explanation of the observed X-ray emission.
In the X-ray reflection nebula (XRN) model, K-shell photo-ionization and the subsequent fluorescence produce a strong Fe K$\alpha$ line with an equivalent width EW$\ge 1$~keV, while a competition between Compton scattering of high energy photons and photoelectric absorption of low energy photons gives rise to a Compton reflection hump around $20-30$~keV \citep{Sunyaev1993, Sunyaev1998, Koyama1996}.
An embedded or nearby transient X-ray source was ruled out to be an illuminating source, since no transient source has been sufficiently bright ($\rm L\ge10^{37}~erg~s^{-1}$) for several years since 1993 within or close to \sgrb\ (e.g. \citealp{Revnivtsev2004, Terrier2010}).
Sgr~A$^\star$ was proposed to be the likely source illuminating \sgrb.
A major outburst from Sgr~A$^\star$, with a luminosity of a few $10^{39}~\rm erg~s^{-1}$ lasting more than 10 years and ending a few hundred years ago would explain the Sgr~B2 emission \citep{Koyama1996, Terrier2010}.
This hypothesis is reinforced by the discovery of a superluminal Fe K echo from the ``Bridge", which points to propagation of an event far away from the clouds \citep{Ponti2010}. 
However, the story gets complicated by different Fe K$\alpha$ line variability detected from GCMCs, which cannot be explained with a single outburst from Sgr~A$^\star$.
The strong and fast variation of the Fe K$\alpha$ line flux in the ``Bridge" region requires a two-year peaked outburst with luminosity of at least $10^{39}\rm~erg~s^{-1}$, while the slower Fe K$\alpha$ line behavior detected in other clouds suggests a second flare with a longer duration \citep{Clavel2013}.

The propagation of cosmic-ray (CR) particles within the molecular clouds is an alternative explanation.
Low-energy cosmic ray electrons (LECRe) and protons (LECRp) can both produce hard X-rays and Fe K$\alpha$ emission \citep{Valinia2000, Dogiel2009}.
The \sgrb\ Fe K$\alpha$ flux variation time scale of $\sim10$~years is too short compared to the Coulomb cooling time of $\sim100\rm~MeV$ protons invoked in the LECRp scenario, thus rules it out as a major contributor to the fast changing Fe K$\alpha$ emission of Sgr~B2 \citep{Terrier2010}.
However, LECRp could be a major contributor to the constant background level of the Sgr~B2 Fe K$\alpha$ emission, which could be detectable once the reflected X-ray emission completely fades \citep{Dogiel2009}.
\citet{Dogiel2009} estimated that LECRp could contribute to about 15\% of the observed maximum Fe K$\alpha$ flux obtained around 2000. 
In the LECRe scenario, a fast variation can be reproduced.
Even if LECRe seems to successfully explain the X-ray emission from several GC molecular clouds \citep{Yusef2007}, it meets challenges for most of the Fe K$\alpha$ bright structures \citep{Dogiel2013, Dogiel2014}.
In the case of \sgrb, the cosmic ray electron energy required to produce the cloud X-ray emission is as high as the bolometric luminosity of the entire cloud \citep{Revnivtsev2004}.  
The derived metallicity of $Z/Z_{\sun}\sim 3.1$ is also much higher than current measurements of the GC metallicity, which ranges from slightly higher than solar \citep{Cunha2008, Davies2009} to twice solar \citep{Giveon2002}.
However, even if LECRe is not the dominant process in the GC \citep{Dogiel2013} and particularly in Sgr B2, we cannot exclude that, in specific regions, the LECRe process contributes to the background level of the Fe K$\alpha$ emission.

Though the GCMCs have been studied extensively below 10~keV, the investigation of the continuum emission extending beyond 10~keV has been limited.
The hard X-rays from the \sgrb\ region were first detected by {\it GRANAT}/ART-P \citep{Sunyaev1993} and then by \integral\ in 2004 \citep{Revnivtsev2004}.
In the XRN scenario, \citet{Terrier2010} derived an illuminating source spectral index of $\Gamma \sim2$ with combined \xmm\ and \integral\ spectra.
Years of monitoring with \integral\ reveal that the hard X-ray emission decreases by a factor of 0.4 from 2003 to 2009 \citep{Terrier2010}.
Nevertheless, \integral\ was not able to resolve the hard X-ray emission.
The accuracy of the hard X-ray luminosity was limited by the unknown distribution of the emission. 

Hard X-ray observations are crucial to better constrain the origin of the GCMC non-thermal X-ray emission.
With unprecedented spatial and spectral resolution in the 10--79~keV band, \nustar\ observed GCMCs in 2013 as part of the Galactic plane survey campaign, including the \sgrb\ region.
For the first time, \nustar\ resolved its hard X-ray morphology and obtained a broadband spectrum from a single instrument.
In Section 2, we introduce the observations and data reduction for \nustar\ and \xmm\ data used for the following analysis.
We present the morphology of the central 90\asec\ radius region of \sgrb\ and the newly discovered cloud feature G0.66$-$0.13 in Section 3, their time variability in Section 4 and their spectroscopy in Section 5.
Based on these observational results, we discuss their implication on both XRN and LECR scenarios in Section 6.

\section{Observation and Data Reduction}

\subsection{\nustar\ Data}

The \nustar\ observatory operates in the broad X-ray energy band from 3 to 79 keV \citep{Harrison2013}. 
\sgrb\ was observed by \nustar\ in October 2013 in two 25\% overlapping pointings, with a total exposure time of 293.7~ks (See Table \ref{tab:obs}).  
 
In both observations, the \sgrb\ region was imaged with the two co-aligned X-ray telescopes, with corresponding focal plane modules FPMA and FPMB, providing an angular resolution of $58\asec$ Half Power Diameter (HPD) and $18\asec$ Full Width Half Maximum (FWHM) over the 3--79~keV X-ray band, with a characteristic spectral resolution of 400~eV (FWHM) at 10~keV. 
The nominal reconstructed \nustar\ astrometry is accurate to $8\asec$ (90\% confidence level, \citealp{Harrison2013}).
The data were reduced and analyzed using the \nustar\ Data Analysis Software NuSTARDAS v.~1.3.1. and HEASOFT v.~6.13, then filtered for periods of high instrumental background due to South Atlantic Anomaly (SAA) passages and known bad/noisy detector pixels. 

The \nustar\ detectors are not completely shielded from incident X-rays that do not go through the optics, which is referred to as stray light. 
Bright X-ray sources within $\sim1$--5 degrees of the \nustar\ field of view can significantly contaminate one or both of the detector planes.
The contaminated detector pixels can be removed based on a numerical model that fully takes into account the telescope geometry \citep{Krivonos2014}.
Thus, to make a stray-light-free mosaic, we used this model to remove stray-light patterns from both FPMA and FPMB detectors by flagging the contaminated detector pixels as bad, when processing with the  {\it NuSTARDAS} pipeline.
As a result, stray-light from the the X-ray sources SLX~1744$-$299 and 1E~1740.7$-$2942 were removed from FPMA detectors, 
and that from GX~3+1 and SLX~1735$-$269 were removed from FPMB detectors.  
We then registered these stray-light-free images with the brightest point sources available in individual observations.
As a result, the astrometry is improved to $\sim 4 \asec$.
Lastly we combined the exposure-corrected images in different energy bands (Figure 1).

On the other hand, removing the stray-light patterns is not necessary for spectral extraction. 
The stray-light background does not significantly change from one sky pointing to another, provided that they are separated by no more than 10\amin--20\amin\ \citep{Krivonos2014}.
Therefore, for the \sgrb\ core region detected in one observation, we extracted the background from the same detector region in the other observation.
This was our motivation for using two 25\% overlapping sky pointings.
This background subtraction method has been applied to many sources detected by \nustar\ suffering from stray-light (e.g. \citealp{Zhang2014, Krivonos2014}), and is proven to subtract mild stray-light contamination properly.
 
Due to the extremely bright stray-light contamination within 90\asec\ of \sgrb\  on FPMB, we used FPMA data only for the \sgrb\ core spectrum.
From the selected data sets, we extracted source spectra from a circular region of 90\asec\ radius centered on Sgr~B2 (R.A.=$17^{h}47^{m}20.4^{s}$, Decl.=$-28^{h}23^{m}07.0^{s}$, J2000).
In this way, we had two source-background pairs with the \sgrb\ core detected on FPMA in observation 40012018002 and 40012019001.
We combined the two FPMA source spectra and their associated response files and background spectra.
The resultant spectrum was grouped such that the detection significance in each data bin is at least $3\sigma$.
The confidence level for all the error bars reported in this paper are 90\%.

Another cloud feature, G0.66$-$0.13 (R.A.=$17^{h}47^{m}41.5^{s}$, Decl.=$-28^{h}26^{m}23.0^{s}$, J2000) \citep{Ponti2014}, is only captured in the second observation (obsID 4001201901) and avoids the stray light from GX~3+1 on FPMB.
We thus used both FPMA and FPMB from the second observation for its spectral analysis.
The spectra were combined and grouped with the same method as for the central 90\asec\ region.

\subsection{\xmm\ Data}

We collected and analyzed all the \xmm\ data available in the archive covering the Sgr~B2 and G0.66$-$0.13 regions.
This includes observations performed in 2000, 2001, 2002, 2004 and 2012.  
The 2000 data were excluded due to low effective exposure and poor statistics, while for the other years we used all the available observations.
The list of \xmm\ observations is presented in Table 1, along with the total EPIC pn-equivalent exposure times (i.e., computed assuming a 0.4 ratio between the effective areas of the MOS and pn detectors).
For each selected observation we extracted the spectra from all available EPIC instruments using the \xmm\ Extended Source Analysis Software  (ESAS; \citealt{Snowden2008}) distributed with version 12.0.1 of the \xmm\ Science Analysis Software. 
For each exposure, calibrated event files were produced with the tasks \textit{epchain} and \textit{emchain} and filtered with \textit{pn-filter} and \textit{mos-filter} in order to exclude the time intervals affected by soft proton contamination. 
The spectra were then extracted with the ESAS \textit{mos-spectra} and \textit{pn-spectra} scripts and rebinned to have at least 30 counts in each bin to apply chi-square statistics. 
The quiescent component of the EPIC internal particle background (QPB) was estimated using archival observations provided within the ESAS database and taken 
with the filter wheel closed.

\begin{deluxetable}{lccccc}[h]                                                                                                                
\tablecaption{\nustar\ and \xmm\ observations of \sgrb.}
\tablewidth{0pt}
\tablecolumns{4}                                                                                                                    
\tablehead{ \colhead{Instrument} & \colhead{Observation}   &   \colhead{Start Time}   &   \colhead{Exposure}   \\
\colhead{ }  & \colhead{ID}  & \colhead{(UTC)}  &  \colhead{(ks)}  }  
\startdata
\nustar\  & 40012018002  & 2013-10-22T16:56:07  & 142.2\\
\nustar\  & 40012019001  & 2013-10-25T22:31:07  & 151.6\\
\hline
\xmm\    &  0112971501   & 2001-04-01T00:25:11  &  9.2\\
\xmm     &  0030540101   & 2002-09-09T11:11:26  &  19.3\\
\xmm     &  0203930101   & 2004-09-04T02:53:45  &  48.5\\
\xmm     &  0694640601   & 2012-09-06T10:56:15 &  66.6\\
\xmm     &  0694641301   & 2012-09-16T18:34:18 &  72.9\\
\xmm     &  0694641401   & 2012-09-30T19:39:50 &  ~~~\;$58.3^{*}$
\enddata
\tablecomments{$^{*}$In the observation 0694641401, the exposure time is 58.3~ks for \sgrb\ and 44.8~ks for G0.66-0.13, as for the latter there are no MOS1 data available.}
\label{tab:obs}
\end{deluxetable}

\section{Spatial Distribution of The Hard X-ray Emission and the Fe K$\alpha$ Line Emission}
 
Figure 1 shows the resultant $17\amin \times 11\amin$ \nustar\ sky mosaics of the \sgrb\ region in the 3--40~keV, 3--10~keV, 6.2--6.6~keV and 10--40~keV bands.
The 3--40~keV image shows that two features are significantly detected: the central 90\asec\ of \sgrb\ and a newly discovered cloud feature G0.66$-$0.13, whose 6.4~keV Fe $\rm K\alpha$ emission turned bright in 2012 as revealed by the \xmm\ data (Terrier et~al. in prep.).
The green circle of 90\asec\ radius outlines the central region of \sgrb, corresponding to $3.4\pm0.3$~pc with the cloud distance of $7.9\pm0.8$~kpc \citep{Reid2009}. 
The green ellipse outlines the cloud feature G0.66$-$0.13, with a semi-major axis of $130\asec$ ($4.9\pm0.5$~pc) and a semi-minor axis of $76\asec$ ($2.9\pm0.3$~pc). 
G0.66$-$0.13 is $\sim14$~pc away from the center of \sgrb\ in the projected plane.
Sgr~B2 and G0.66$-$0.13 are both about 100~pc away from Sgr~A$^{\star}$ in the projected plane.
The lower energy 3--10~keV image is also shown to compare with previous observations of \sgrb\ by \chandra, \XMM, and other imaging observatories. 
The 6.2--6.6~keV band image with continuum emission subtracted shows the Fe $\rm K\alpha$ line emission morphology of the \sgrb\ region.
The 10--40~keV band image provides the line-free continuum emission morphology.
All the images are scaled individually to illustrate the morphology of major features. 
The images are overlaid with 6.4~keV line emission contours.
The contours are made from the 2012 image of the \xmm\ CMZ scan in the 6.28--6.53 keV band, from which the continuum emission (estimated assuming a power law model between two adjacent bands) has been subtracted (\citealp{Ponti2014}, 2015 in press).

The 10--40~keV image demonstrates that the high energy X-ray ($\ge10$~keV) morphology of \sgrb\ is resolved at sub-arcminute scales for the first time.
It clearly reveals substructures of \sgrb\ and proves that the X-ray emission in this energy band is extended (Figure~1).
Both the central 90\asec\ of \sgrb\ and G0.66$-$0.13 are significantly detected above 10~keV.
G0.66$-$0.13 shows two bright cores separated by about 100\asec, well correlated with its 6.4~keV line contour.
The size of each peak is $\sim 20$\asec. 
The 10--40~keV surface brightness of the whole G0.66$-$0.13 region is $(3.5\pm0.8) \times 10^{-6} \rm ~ph~cm^{-2}~s^{-1}~arcmin^{-2}$.
Within the central 90\asec\ region of \sgrb, the 10--40~keV emission peaks at the center, coinciding with the compact star-forming core Sgr B2(M), with a surface brightness of $(1.6\pm0.1)\times 10^{-5}\rm~ph~cm^{-2}~s^{-1}~arcmin^{-2}$ for the central 25\asec\ radius region.
It is detected at the $\sim17\sigma$ level and likely to be the hard X-ray counterpart of Sgr~B2(M).
The right panel of Figure~2 shows the zoomed-in $7\amin \times 5\amin$ image of the central region in 10--40~keV. 
Besides Sgr~B2(M), X-ray emission around the additional compact core Sgr~B2(N), about 50\asec\ north of Sgr~B2(M), is also detected ($\sim5\sigma$). 
The zoomed-in images in Figure~2 are overlaid with the source regions of Sgr~B2(M) (RA$=17^{h}47^{m}20.30^{s}$, Dec$=-28^{\circ}23\amin04.01\asec$, J2000) and Sgr~B2(N) (RA$=17^{h}47^{m}20.30^{s}$, Dec$=-28^{\circ}23\amin04.01\asec$, J2000) defined from submm band observations by {\it Herschel} \citep{Etxaluze2013}.
X-ray emission from a third compact core Sgr~B2(S), south of Sgr~B2(M), is not detected in 10--40~keV.
The surrounding regions show lower surface brightness, with the western half of the annulus from 25\asec\ to 90\asec\ brighter than the eastern half. 
The 10--40~keV morphology strongly resembles the optical depth map at 250~$\mu$m, which indicates the local column density, derived from submm continuum emission \citep{Etxaluze2013}.
In the 250~$\mu$m optical depth map, Sgr~B2(M) and Sgr~B2(N) have highest optical depth at 250~$\mu$m ($\tau_{250\mu m}>1$), while the surrounding areas have gradually lower $\tau_{250\mu m}$ with the western region higher than the eastern.  
It suggests that the hard X-ray continuum emission traces the local column density of the cloud material.

In 6.2--6.6 keV, only the main compact core Sgr~B2(M) is significantly detected.
Sgr~B2(N) is not detected at 6.4~keV, nor in the 3--10~keV range (left panel of Figure 2).
This could be due to higher local absorption.
G0.66$-$0.13 is also not detected by \nustar\ at 6.4~keV. 
This is a dramatic change from its 2012 Fe K$\alpha$ line morphology represented by the cyan contours, where G0.66$-$0.13 was brighter than the \sgrb\ core region.
This indicates that the G0.66$-$0.13 Fe K$\alpha$ emission has a short life time (e-folding decay time) of about 1 year.

Finally, we note that the small white circle with 16\asec\ radius illustrates a bright point source CXOUGC J174652.9$-$282607, which is registered in the \chandra\ point source catalogue \citep{Muno2009}.
The source is detected up to $\sim40$~keV by \nustar\ (R.A.$=17^{h}47^{m}52.968^{s}$, Decl.$=-28^{h}26^{m}07.37^{s}$, J2000), and will be reported in another work (Hong et~al. in prep.).

\begin{figure*} 
\centering
\label{fig:mosaic}
\begin{tabular}{cc}
\includegraphics[height=0.3\linewidth]{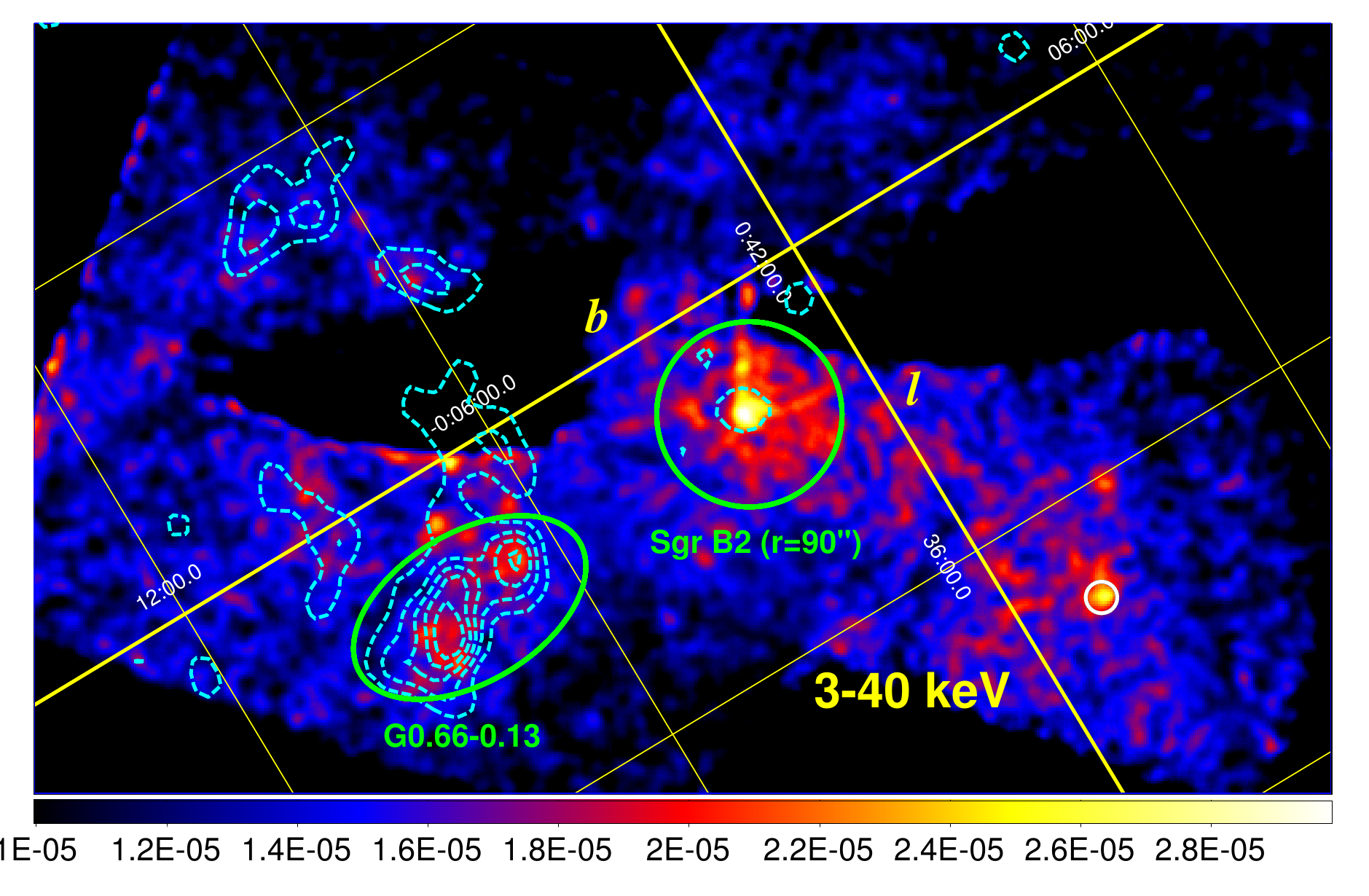} &
\includegraphics[height=0.3\linewidth]{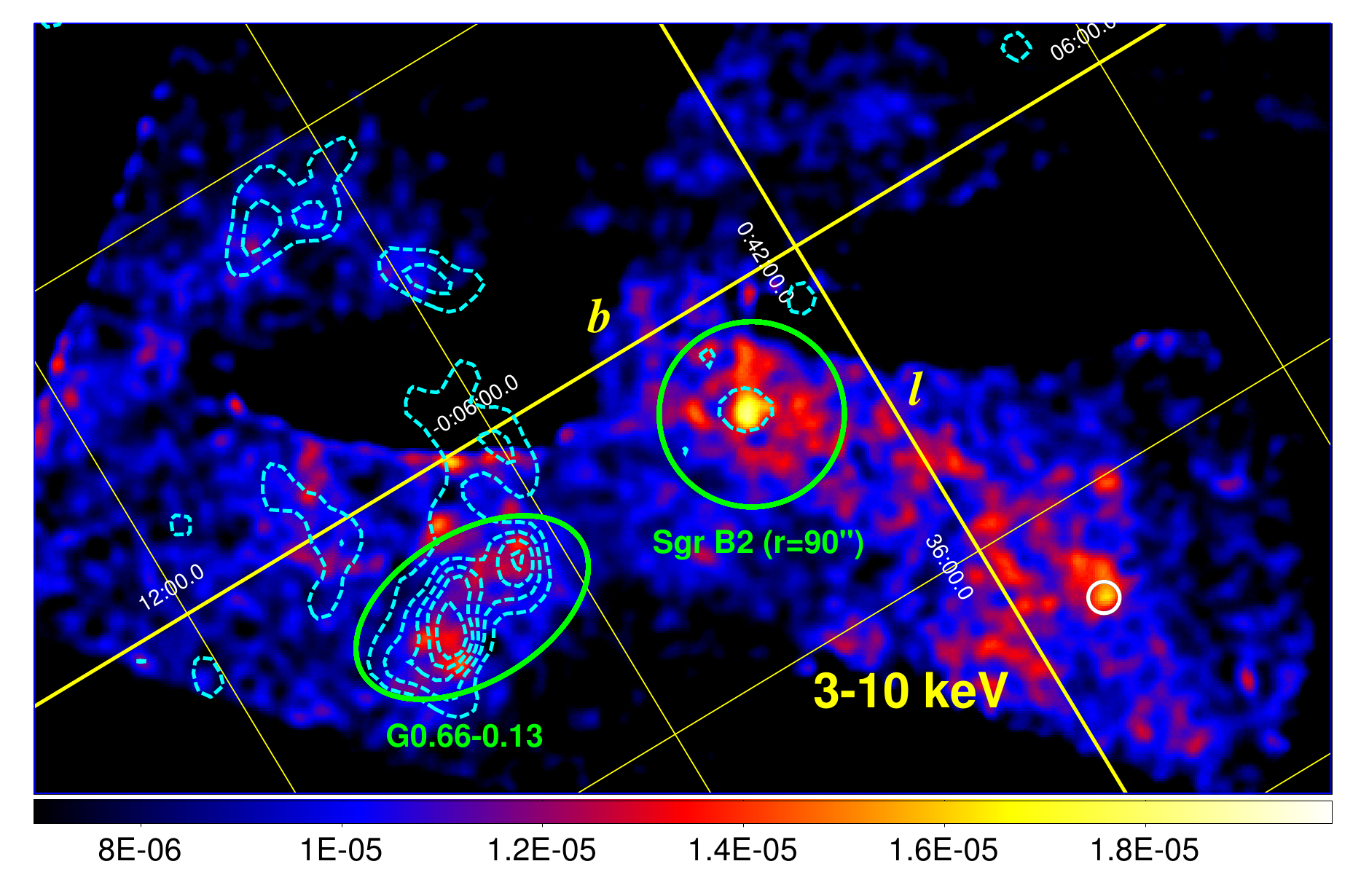} \\
\includegraphics[height=0.3\linewidth]{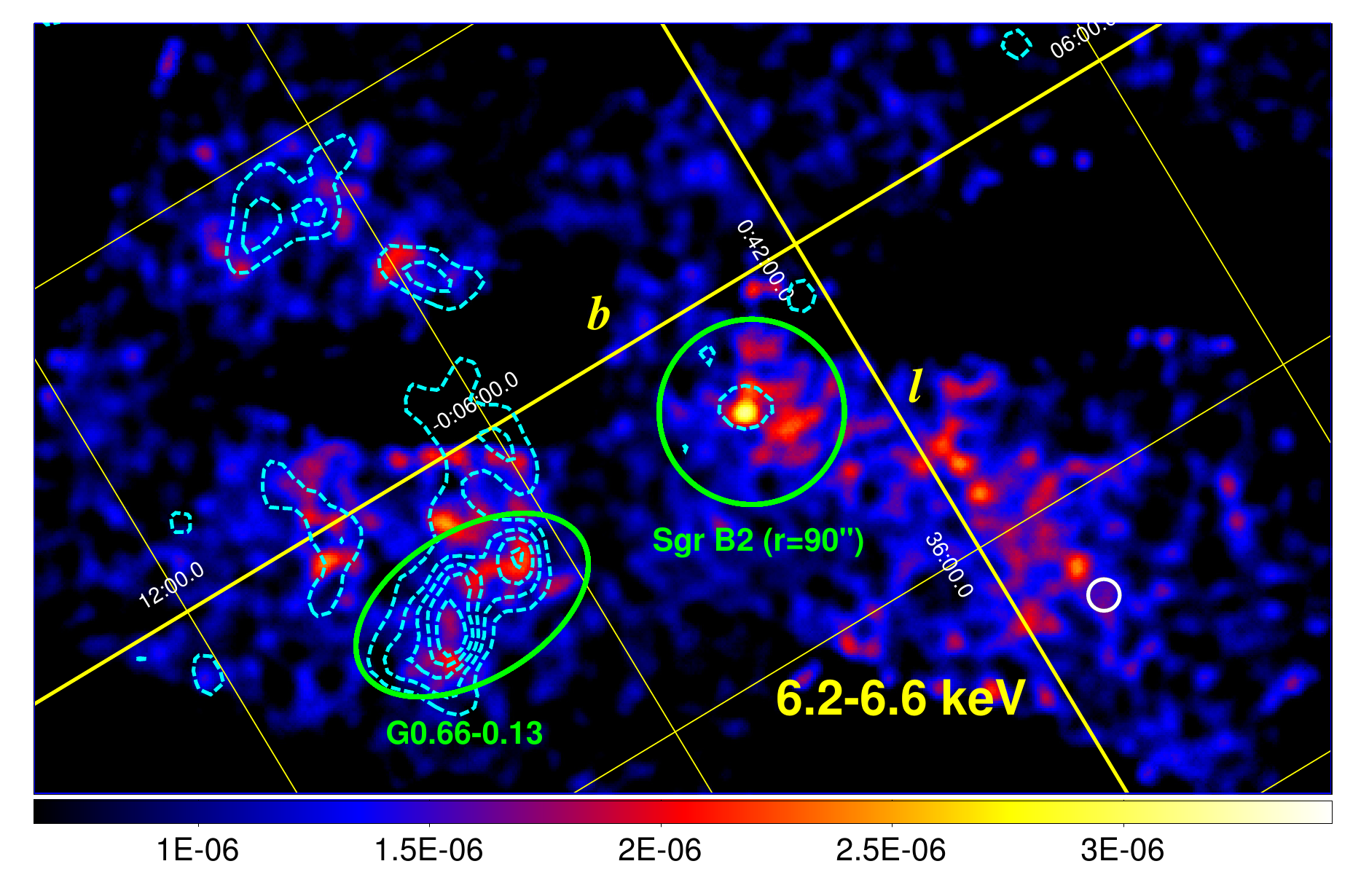} &
\includegraphics[height=0.3\linewidth]{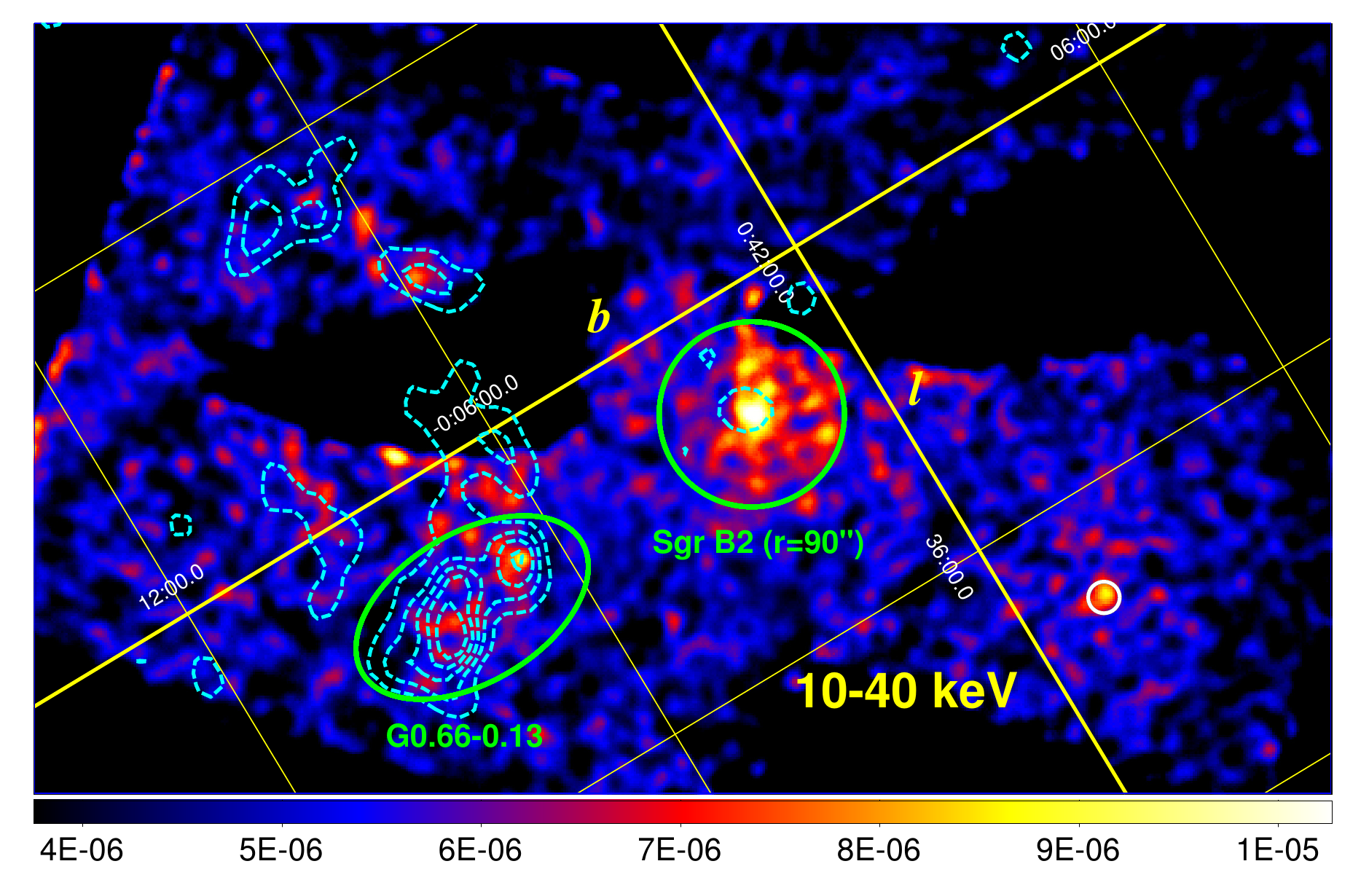} \\
\end{tabular}
\caption{
Upper left panel: 17\amin $\times$11\amin\ \nustar\ 3--40~keV mosaic image of the \sgrb\ region overlaid with 6.4 keV line emission contours (cyan dashed) made from 2012 \xmm\ observations of the same region. The green circle is the central 90\asec\ of \sgrb\ and the green ellipse shows the new cloud feature G0.66$-$0.13. Both the Sgr B2 core and G0.66$-$0.13 are detected in 3--40~keV. Upper right panel: \nustar\ 3--10~keV mosaic image of the \sgrb\ region.The small white circle shows a bright point source CXOUGC J174652.9$-$282607. Lower left panel: \nustar\ 6.2--6.6~keV mosaic image of the \sgrb\ core region from which the continuum emission is subtracted. The inner 90\asec\ of \sgrb\ correlates well with its 2012 emission. While G0.66$-$0.13 was brighter than the inner 90\asec\ of \sgrb\ in 2012, it is not detected in the 6.2--6.6~keV energy band by \nustar\ in 2013. Lower right panel: \nustar\ 10--40~keV mosaic image. The central 90\asec\ of \sgrb\ and G0.66$-$0.13 are both detected in 10--40~keV. All the images are in $\rm cts~s^{-1}$ and overlaid with the Galactic coordinates with a grid of $0.1^{\circ}$. Sgr~A$^\star$ is outside of the field of view to the bottom right. The dark regions with irregular shapes are those of no exposure, which are cut out due to contamination of stray-light from bright point sources outside the field-of-view.  
}
\end{figure*}

\begin{figure*} 
\centering
\label{fig:core}
\begin{tabular}{cc}
\includegraphics[height=0.3\linewidth]{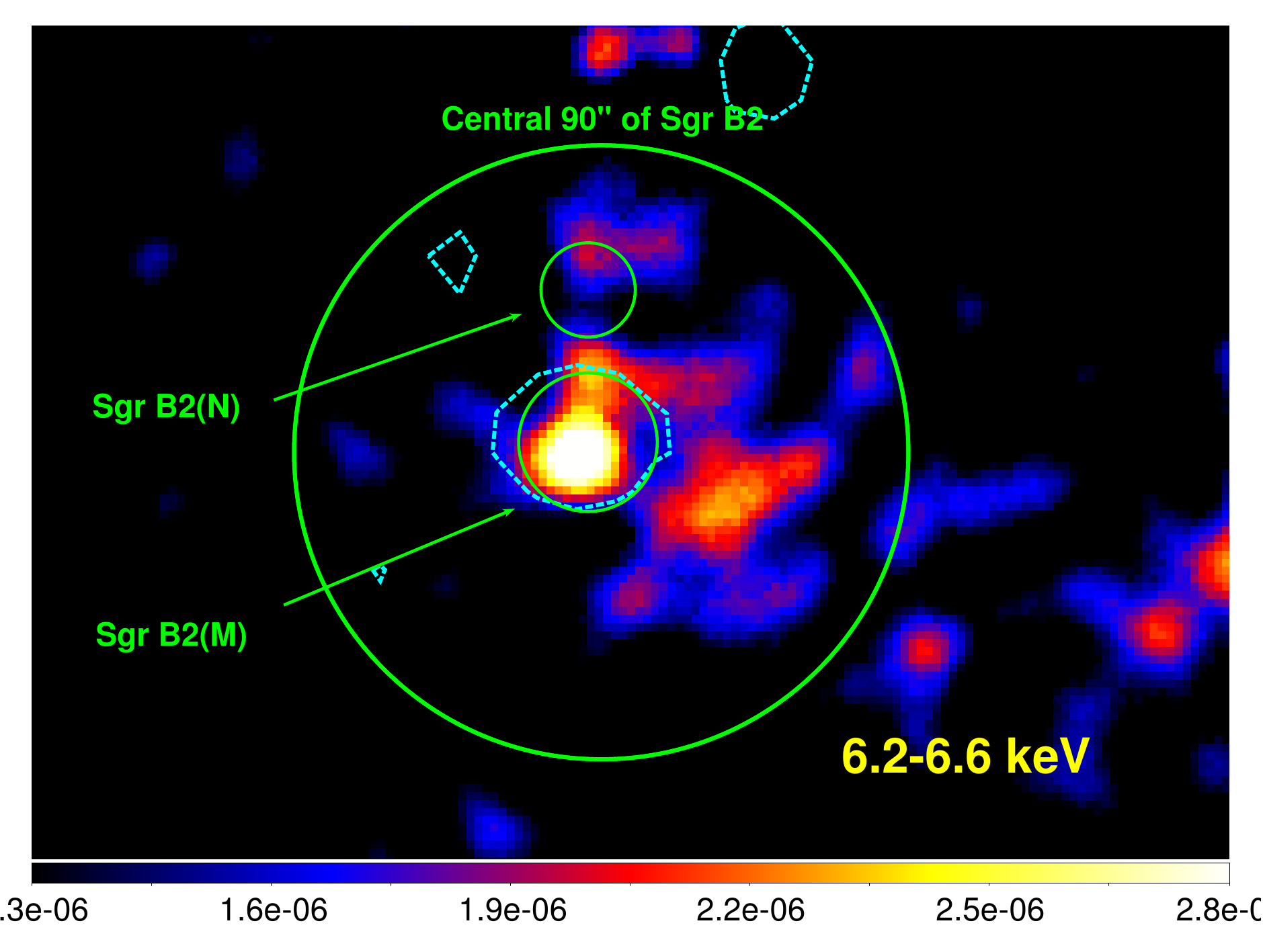} &
\includegraphics[height=0.3\linewidth]{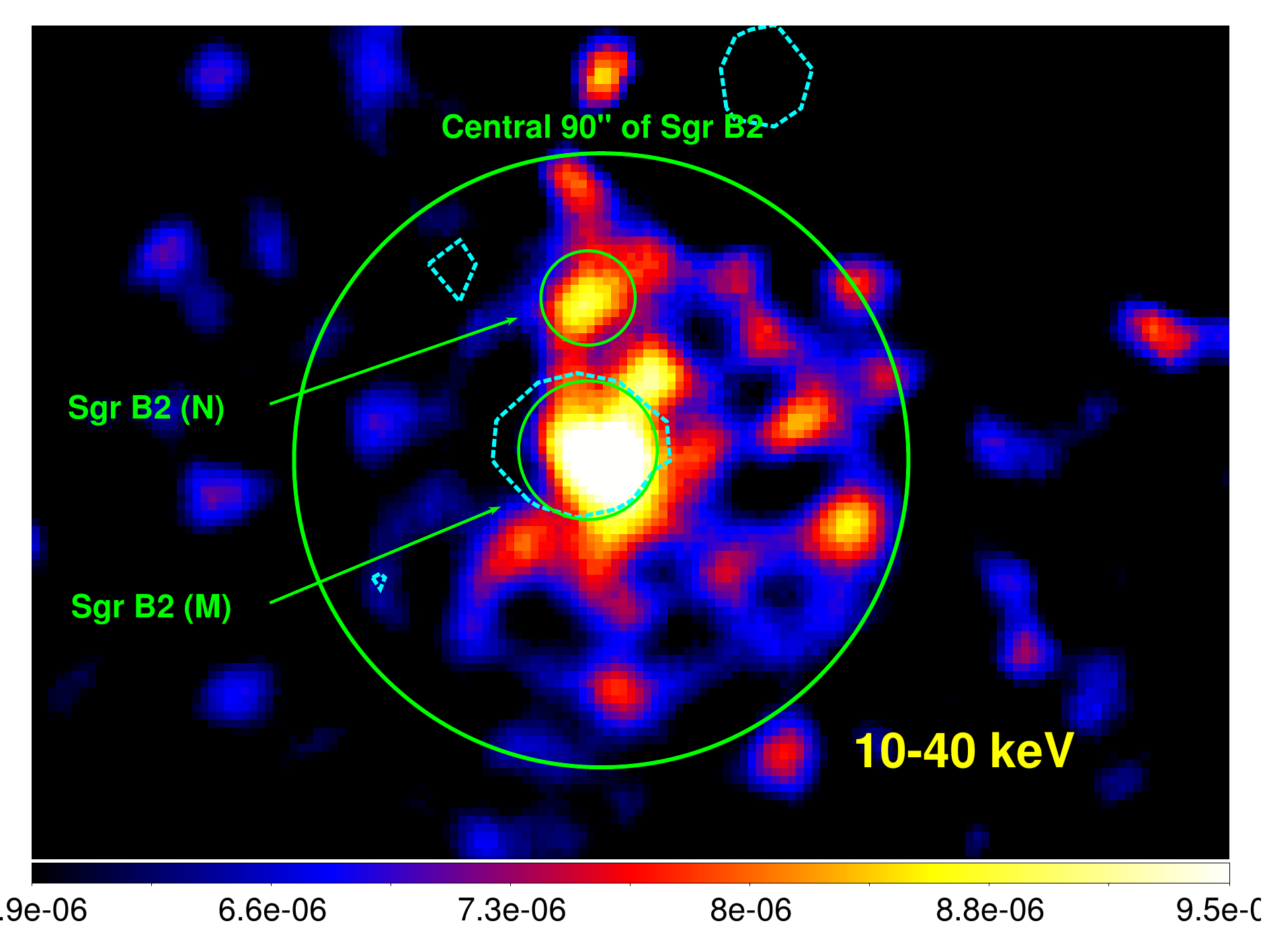} \\
\end{tabular}
\caption{Zoomed-in 6\amin $\times$ 4\amin\ \nustar\ mosaic image of the central \sgrb\ region in 6.2--6.6~keV with continuum emission subtracted (left panel) and 10--40~keV (right panel). Both images are in $\rm cts~s^{-1}$ and overlaid with 6.4 keV line emission contours (cyan dashed). The two main compact cores Sgr~B2(N) and Sgr~B2(M) are outlined with 15\asec\ radius region and 20\asec\ radius region, respectively. The region for the central 90\asec\ of \sgrb\ is overlaid for comparison. Sgr~B2(M) is detected in both energy bands, while Sgr~B2(N) is only detected above 10~keV.   
}
\end{figure*}


\section{Time Variability of Central \sgrb\ and G0.66$-$0.13}

With the \xmm\ data in 2001, 2002, 2004 and 2012 and the most recent 2013 \nustar\ observation of \sgrb, we obtained the observed 6.4~keV line flux for the central 90\asec\ radius region of \sgrb\ using a simple model {\tt wabs*(apec+wabs*pow+gauss+gauss)} (see Section 5.1.1) as the observed line flux is model independent.
All the model parameters are fixed to be the same for all the years, while the normalizations for the power-law and Gaussian components are left free.
The fit is satisfactory with $\chi^2_{\rm \nu}=1.0$ for $d.o.f.=517$.
The resultant absorbed flux change of the central 90\asec\ of \sgrb\ over time is shown in Figure 3 with square data points.
The 6.4 keV line flux shows a clear decreasing trend from 2001 to 2013, which can be fit with a linear decay model with a slope of $s=-0.25\times10^{-5}\rm~ph~cm^{-2}~s^{-1}~yr^{-1}$.
The varying flux is preferred at a $\sim7\sigma$ level over a flat light curve.
In 2013 the observed 6.4~keV line flux is down to $(0.83\pm0.21) \times 10^{-5} \rm~ph~cm^{-2}~s^{-1}$, which corresponds to about 20\% of the 2001 observed flux.
The 6.4 keV line flux decrease rate is compatible with previous works showing that the 6.4 keV line emission of a more extended Sgr~B2 region in 2005 is about 60\% that of 2000 \citep{Inui2009}.
The life time of the X-ray photons is $t \sim11$~years, which agrees well with the $\sim11$-year light crossing time for the central 90\asec, with a distance of 7.9~kpc \citep{Reid2009}.
Although the 2001-2013 lightcurve shows a decreasing trend, we note that the 2013 Fe K$\alpha$ emission is at the same level as that of 2012.
With limited statistics, it is not conclusive whether the Sgr~B2 Fe K$\alpha$ emission was still decreasing in 2013 or it had reached a constant background emission level.
Future observation of the Sgr~B2 region will constrain its timing variability.

Seven-year monitoring of \sgrb\ by \integral\ reveals that the hard X-ray continuum emission flux decreases with a similar trend to the 6.4~keV emission, both decaying up to 40\% from 2003 to 2009 \citep{Terrier2010}. 
This decay profile predicts that the hard X-ray emission in 2013 reaches $\sim 30\%$ that of 2001.  
The 10--40~keV flux of the central 90\asec\ of \sgrb\ measured by \nustar\ in 2013 is $F_{10-40\rm~keV}=(1.9\pm0.2) \times 10^{-12} \rm~erg~cm^{-2}~s^{-1}$.
Therefore, the extrapolated 10--40~keV flux of the central \sgrb\ region in 2001 is $F_{10-40\rm~keV} \sim 6.3 \times 10^{-12} \rm~erg~cm^{-2}~s^{-1}$ based on the hard X-ray decay profile.  
To verify the estimated flux in 2001, we extrapolated the 2001 \xmm\ spectrum into higher energies and derived the 10--40~keV flux of the central 90\asec\ in 2001 to be $F_{10-40\rm~keV}=(6\pm2) \times 10^{-12} \rm~erg~cm^{-2}~s^{-1}$, which matches with the 2001 flux value extrapolated based on \nustar\ measurements.  
This will be used to derive the luminosity of the primary source in Section 6.1.

The newly discovered cloud feature G0.66$-$0.13, 6\amin\ (14~pc) from the central \sgrb\ region in the projected plane, was not significantly brighter than surrounding regions at 6.4~keV until 2012.  
We derived its 6.4~keV line flux from \xmm\ observations in 2001, 2004 and 2012 (stars in Figure~3). 
The 2002 \xmm\ observation has very poor statistics for this region and thus was not used.
The 2001 flux is poorly constrained to $F_{6.4\rm~keV}=0.05\times10^{-5}\rm~ph~cm^{-2}~s^{-1}$ with an upper limit of $1.60\times10^{-5}\rm~ph~cm^{-2}~s^{-1}$.
The 2001 and 2004 Fe K$\alpha$ emission of G0.66$-$0.13 remained at the same level as the surrounding regions.
The 2012 \xmm\ observation of \sgrb\ clearly revealed a significantly increased 6.4~keV flux, twice that of 2004 and higher than the central \sgrb\ area.
The 6.4~keV line emission contours shown in Figure~1 also shows that the brightest subregions within G0.66$-$0.13 were brighter than the \sgrb\ core at 6.4~keV in 2012.
The G0.66$-$0.13 Fe K$\alpha$ emission experienced an increase prior to 2012 and a fast decay from 2012 to 2013.
However, the 6.4 keV line flux sharply decreases and results in a non-detection by \nustar\ in 2013, giving an upper limit (90\% confidence level) to the 6.4~keV line flux of $5\times10^{-6}\rm~ph~cm^{-2}~s^{-1}$.
The 6.4~keV line flux in 2013 is less than 50\% of the value measured by \xmm\ in 2012.
It requires a short lifetime of $\sim1$~year, which roughly matches the light crossing time of the two bright cores within G0.66$-$0.13 (2-3~years). 

\begin{figure}
\centeringÊ 
\label{fig:time_variability}
\begin{tabular}{cc}
\includegraphics[angle=0, width=0.45\textwidth]{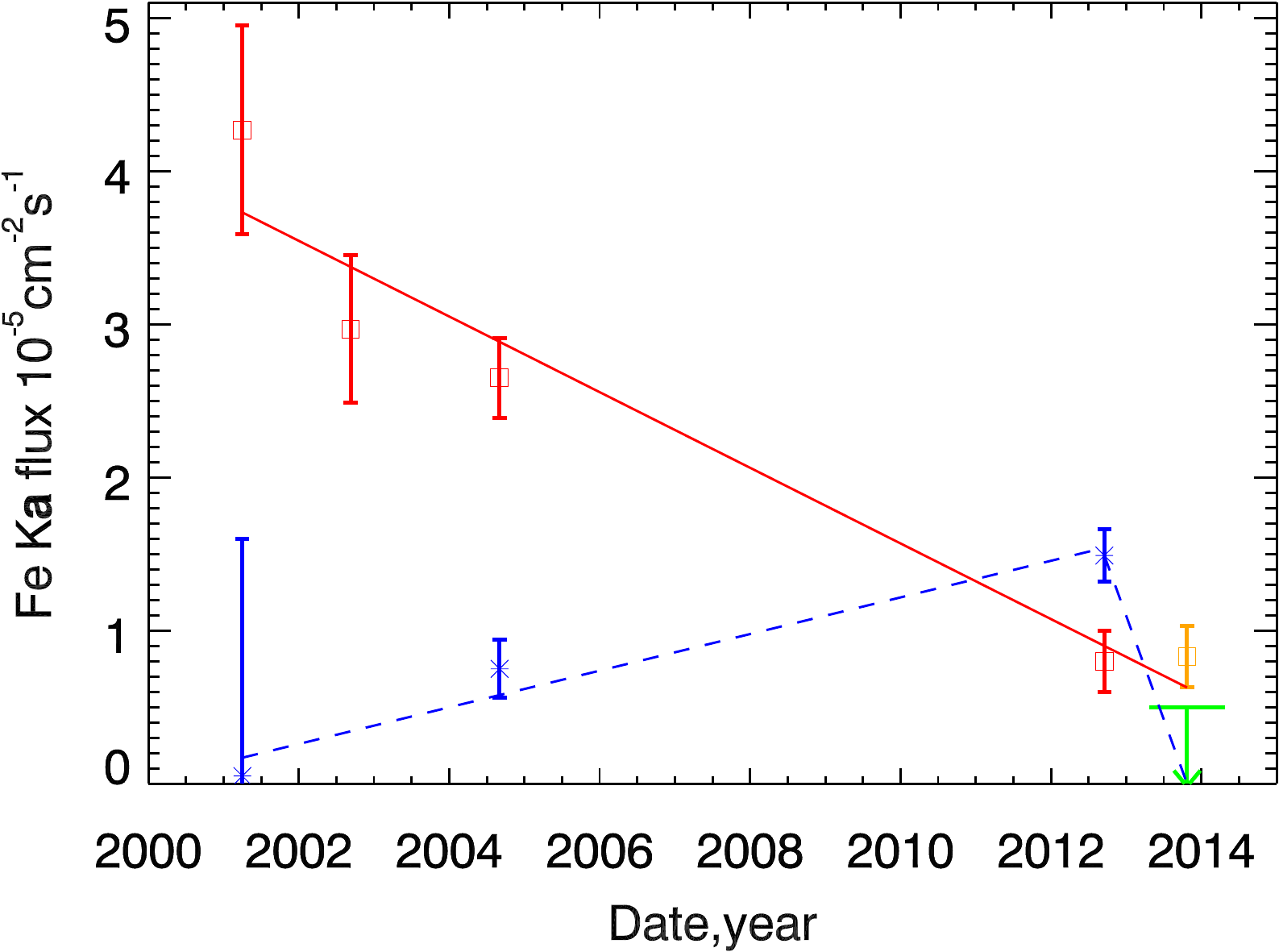} 
\end{tabular}
\caption{Time variability of the Fe K$\alpha$ line for the central 90\asec\ of \sgrb\ and the additional cloud feature G0.66$-$0.13. The \sgrb\ absorbed 6.4~keV line flux was measured in 2001, 2002, 2004 and 2012 by \xmm\ (red squares), and in 2013 by \nustar\ (orange square). The Fe K$\alpha$ line flux shows a clear decay of up to 80\% of the measured \sgrb\ flux from 2001 to 2013. A chi-square hypothesis test favors a linearly decreasing flux at a 6.7$\sigma$ level, the red line shows the best-fit linearly decay model with a slope of $s=-0.25\times10^{-5}\rm~ph~cm^{-2}~s^{-1}~yr^{-1}$. The absorbed 6.4~keV line flux of G0.66$-$0.13 was measured in 2001, 2004 and 2012 by \xmm\ (blue stars) and in 2013 by \nustar\ (green upper limit). The G0.66$-$0.13 Fe K$\alpha$ emission experienced an increase before 2012 and a fast decay from 2012 to 2013. The blue line shows the best-fit linearly increasing model with a slope of 0.12. Yet in 2013 the 6.4 keV line was not significantly detected by \nustar.}
\end{figure}

\section{Broadband 2013 Spectra of Central SGR~B2 and G0.66$-$0.13}

\subsection{Central 90\asec\ Radius Region of Sgr~B2}

We extracted a \nustar\ spectrum from the central 90\asec\ radius region of Sgr~B2 (R.A.=$17^{h}47^{m}20.4^{s}$, Decl.=$-28^{h}23^{m}07.0^{s}$, J2000).
The background subtraction method leaves mainly the molecular cloud emission and previously detected X-ray point sources.
Thus, we first checked the flux contribution from known point sources in the source and background regions. 
The background regions we use do not contain any point sources above the \chandra\ detection threshold.
In the source region, there are four point sources detected by \chandra\ within 90\asec\ of \sgrb : CXOUGC~J174723.0$-$282231, CXOUGC J174720.1$-$282305, CXOUGC~J174718.2$-$282348 and CXOUGC~J174713.7$-$282337. 
We checked each source using the stacked spectrum made from all archived \chandra\ data between 1999-09-21 and 2012-10-31 available for \sgrb.
Their summed observed flux corresponds to $\sim 7\%$ of the total observed flux of the central 90\asec\ region below 10~keV.
By extrapolating their spectra into a higher energy band, we estimate their flux contribution above 10~keV is only $\sim 4\%$.
Therefore the flux contribution from point sources is not significant for our study. 
None of the four sources exhibits a 6.4~keV line feature,
while the brightest one among the four point sources exhibits a significant 6.7~keV line feature in its spectrum.
It can be best fit with a collisionally ionized plasma model {\tt apec} with a temperature of $\sim 3$~keV.
We thus used this model to represent unresolved point X-ray sources and any possible residual GC diffuse X-ray background within the source region.

\subsubsection{Ad hoc XRN Model}

To examine the XRN scenario, we first applied a popular XRN model, with the {\tt wabs} absorption model for a direct comparison with previous work.
The model is composed of a power-law and two Gaussians representing the Fe $\rm K\alpha$ emission line at 6.40~keV and the $\rm K\beta$ line at 7.06~keV, modified by intrinsic absorption column density.
All the model components are subject to the foreground absorption, resulting in the {\tt XSPEC} model {\tt wabs*(apec+wabs*powerlaw+gauss+gauss)}. 
We fixed the line energies of the two Gaussians to be 6.40~keV (best-fit centroid energy of the Fe K$\alpha$ line of $\sim6.44$~keV) and 7.06~keV, with the normalization ratio of $\rm K\beta/K\alpha$ set to 15\%  \citep{Murakami2001}. 

The 3--79~keV spectrum is well-fit with this simple model ($\chi^2_{\rm \nu}$=0.97 for $d.o.f.=91$).
The best-fit model parameters are listed in Table~2.
The temperature of the thermal component is $2\pm1$~keV with an unabsorbed $2-10$~keV flux of $F_{2-10\rm~keV} =\rm (6\pm4) \times10^{-13} \rm~erg~cm^{-2}~s^{-1}$, consistent with the spectra of the point sources within 90\asec\ of \sgrb.
The 6.4~keV Fe K$\alpha$ line has an observed flux of $F_{6.4\rm~keV}=\rm (8.3 \pm 2.1) \times 10^{-6}~ph~cm^{-2}~s^{-1}$.
We calculated its equivalent width (EW) based on the power-law as the only continuum component, resulting in EW$=1.2^{+0.7}_{-0.3}$~keV.
The best-fit power-law photon index is $\Gamma=1.9\pm0.5$, consistent with previous measurements of $\Gamma \approx 2$ \citep{Terrier2010}.
The observed 10--40~keV flux is $F_{10-40\rm~keV}=\rm (1.9\pm 0.2)  \times 10^{-12} \rm~erg~cm^{-2}~s^{-1}$.
The intrinsic absorption column density is found to be $N_{\rm H}(i)=\rm (5.0\pm1.3)\times10^{23}~cm^{-2}$, on the lower end of, but still consistent with, the previous result of $\rm (6.8\pm 0.5) \times 10^{23}~cm^{-2}$ derived using the combined 2003 \xmm\ and $2003-2004$ \integral\ data \citep{Terrier2010}.
All components are subject to a foreground interstellar column density of $N_{\rm H}(f)=\rm (1.1\pm0.4)\times10^{23}~cm^{-2}$. 
The foreground column density value is consistent with our analysis of accumulated 2001-2012 \xmm\ data of the inner 90\asec\ of \sgrb, which gives $N_{H} (i)=(1.0 \pm 0.2) \times \rm~10^{23}~cm^{-2}$.
 
Although this ad hoc model can fit well to the data, it is not self-consistent, with the continuum emission and fluorescence lines decoupled. 
A power-law can only measure the spectral slope of the observed scattered continuum, but not the illuminating source spectrum. 
The model is valid for measuring the illuminating source spectrum only when a molecular cloud is optically thin ($N_{H} \ll 10^{24}\rm~cm^{-2}$) and the Compton scattering is negligible.
Interpretation of the resultant best-fit value for the intrinsic column density $N_{H}(i)$ calls for caution.
As this ad hoc model does not take cloud geometry into account, the $N_{H}(i)$ measured by this model represents a characteristic column density of the cloud, while in reality illuminating X-ray photons are absorbed and scattered in various locations of the cloud. 

\subsubsection{Self-consistent XRN Model {\tt MYTorus}}

In the XRN scenario, to consistently measure the illuminating X-ray spectrum and to properly determine the intrinsic column density, a self-consistent XRN model based on Monte-Carlo simulations is required. 
\citet{Murakami2001}, \citet{Revnivtsev2004}, \citet{Terrier2010} and \citet{Odaka2011} have applied Monte-Carlo based XRN models to \sgrb\ data in order to study its morphology and spectrum.
The {\tt MYTorus} model is the only XRN model available in {\tt XSPEC} to self-consistently measure the illuminating source spectrum and the intrinsic column density \citep{Murphy2009}.  
{\tt MYTorus} was originally developed for Compton-thick Active Galactic Nuclei (AGN) assuming a toroidal reflector with neutral materials and uniform density.
The default Fe abundance in the {\tt MYTorus} model is one solar and does not allow variation.
There are three components offered in the model: the transmitted continuum (MYTZ), the scattered continuum (MYTS) and the iron fluorescence lines (MYTL).
In the case of GCMCs, the observed spectrum only contains the last two components, because we are seeing only the reflected X-ray photons off the cloud. 
We thus use the combination of MYTS and MYTL.
Both components depend on the intrinsic equatorial hydrogen column density $N_H(i)$, the illuminating source spectrum photon index $\Gamma$, the inclination angle $\theta_{obs}$ between the line-of-sight and the torus symmetry axis, and the model normalization $N_{MT}$.
To self-consistently measure the illuminating source spectrum, all the parameters are linked between MYTS and MYTL as a coupled mode where the same incident X-ray spectrum is input into both components. 
We select a termination energy of the incident power-law as 500 keV.
As the best-fit energy centroid for the Fe K$\alpha$ line is $\sim 6.44\rm~keV$, we select an energy offset of $+40\rm~eV$ for the MYTL component to allow freedom for the centroid energy. 
The resultant model is {\tt wabs*(apec+MYTS+MYTL)}, where the two {\tt MYTorus} components are implemented as table models in {\tt Xspec}:
{\tt atable\{mytorus\_scatteredH500\_v00.fits\}+ atable\{mytl\_V000010pEp040H500\_v00.fits\}}.

The coupled mode of the {\tt MYTorus} model can be applied to the GCMC X-ray reflection spectra with some restrictions, by treating a quasi-spherical cloud as part of a virtual torus and rescaling incident X-ray flux properly (see Figure~A6 for geometry).
Due to the toroidal geometry and the uniform density profile {\tt MYTorus} assumes, for GCMCs we restrict the applicability of the model to the inclination angle range of $\theta_{obs} \lesssim 60^{\circ}$ and the equatorial column density range of $N_{H} \lesssim 10^{24}\rm~cm^{-2}$, in order to be insensitive to the torus geometry (see Appendix for more details).
When $N_{H}(i)$ reaches $\sim 10^{24}\rm~cm^{-2}$, we derive the systematic error based on the {\tt MYTorus} model to be $25\%$ for $N_{H}$, $3\%$ for $\Gamma$ and $10\%$ for the model normalization (Mori et~al. submitted).
We note that the $N_{H}(i)$ measured by the {\tt MYTorus} model is an averaged value over the torus. 

This self-consistent model can fit to the Sgr~B2 spectrum well, yielding $\chi^2_{\rm \nu}=1.09$ for $d.o.f.=92$ (left panel of Figure~4).
Both the best-fit temperature and the flux of the {\tt apec} component are consistent with the fitting results of the ad-hoc power-law model (Table~2).
The foreground column density is $N_{H}(f)=(1.2 \pm 0.1) \times 10^{23}\rm~cm^{-2}$, consistent with that derived from the ad hoc model and better constrained. 
However, the intrinsic equatorial column density is $N_{H}(i)=(1.01\pm 0.16_{stat} \pm 0.25_{sys}) \times 10^{24}\rm~cm^{-2}$, which is twice the $N_{H}(i)$ value measured with the ad hoc model.
The $N_{H}(i)$ value derived with the two models cannot be directly compared due to the lack of geometry definition for $N_{H}(i)$ in the ad hoc model.
$N_{H}(i)$ in the {\tt MYTorus} model corresponds to the minor diameter of the torus (or diameter of the quasi-spherical cloud), while $N_{H}(i)$ in the ad hoc model roughly corresponds to the cloud radius as it ``measures" an averaged effect, which might explain the difference by a factor of 2 (see the Appendix).
For the central 90\asec\ of \sgrb, the cloud radial optical depth due to Thomson scattering derived by the {\tt MYTorus} model is $\tau_{T}=0.67 \pm 0.27$, consistent with previous measurements of $\tau_{T}=0.4\pm0.1$ derived from a self-developed Monte-Carlo simulation with combined {\it ASCA/\rm GIS}, {\it GRANAT/\rm ART-P} and \integral\ data when \sgrb\ was much brighter \citep{Revnivtsev2004}.   
As a result, the central 90\asec\ of \sgrb\ is marginally optically thin.
The illuminating source spectrum photon index is constrained to be $\Gamma=2.2\pm0.4$ with the systematic error negligible, compatible with the observed spectrum slope of $\Gamma=1.9\pm0.5$ measured by the power-law model which does not consider Compton scattering.
With the hard upper limit on the inclination angle $\theta_{obs}=60^{\circ}$ chosen for this study, the {\tt MYTorus} model is not sensitive to the parameter $\theta_{obs}$. 
Therefore, this parameter could not be constrained by our work.  

Though with limitations, {\tt MYTorus} is currently the best self-consistent XRN model available.
The systematic errors could be underestimated since they are estimated solely based on the {\tt MYTorus} model (see the Appendix).
A modified XRN model based on {\tt MYTorus} with a spherical reflector and variable Fe abundance is under development for the specific case of GCMCs, which will remove the uncertainties induced by the toroidal reflector and the fixed Fe abundance (T. Yaqoob, private communication). 
The more sophisticated XRN model will also implement the specific density profile of Sgr~B2 to remove uncertainties induced by non-uniform density and better constrain $N_{H}(i)$ (Walls et~al. in prep.). 

\subsubsection{Self-consistent Cosmic Ray Electron Model {\tt LECRe}}

To determine whether CR electrons can be a major contributor to the remaining Sgr~B2 X-ray emission in 2013, we tested the {\tt LECRe} model \citep{Tatischeff2012}.
The {\tt LECRe} model has four free parameters: the power-law index of the accelerated particle source spectrum, $s$, the path length of cosmic rays in the non-thermal X-ray production region, $\Lambda$, the minimum electron energy, $E_{min}$, and the model normalization, $N_{LECR}$, which provides the power injected in the interaction region by primary cosmic-ray electrons of kinetic energies between $E_{min}$ and $E_{max}=1\rm~GeV$: $dW/dt=4\pi D^2 N_{LECR}\rm~erg~s^{-1}$, where $D$ is the source distance. 
Since the fit cannot constrain the path length of {\tt LECRe}, we fix it to $\Lambda=5\times10^{24} \rm ~H-atoms~cm^{-2}$ as was used in \citet{Tatischeff2012}.
As shown in the middle panel of Figure~4, it results in an acceptable fit to the data with $\chi^2_{\rm \nu}=1.18$ for $d.o.f.=92$ (Table~2).
The electron spectral index is found to be $s=2.8\pm0.5$ with a normalization of $N_{LECRe}=(2.8\pm0.2)\times10^{-7} \rm ~erg~cm^{-2}~s^{-1}$. 
Previous analysis with \xmm\ and \integral\ data found that $s \sim 1.5$, but with an unconstrained error bar \citep{Terrier2010}.
The best-fit minimum electron energy is as low as 3~keV, with an upper bound of 50 keV.
Electrons with such low energy cannot leave their acceleration site, nor penetrate the cloud. 
The best-fit molecular cloud metallicity is $Z/Z_{\sun}=4.0^{+2.0}_{-0.6}$, higher than $Z/Z_{\sun}=3.1\pm0.2$ derived by \citet{Terrier2010}. 
Such a metallicity value is too high compared to current measurements of the GC interstellar medium metallicity which ranges from slightly higher than solar to twice solar.
Furthermore, compared to the XRN models, the LECRe model results in a poorer fit above 10~keV as it cannot fit to the spectrum curvature equally well.
The high energy part of the spectrum thus provide a more excluding constraint on the {\tt LECRe} model, which does not depend on the high metallicity required by the significant Fe K$\alpha$ line.
We therefore confirm that the unphysical best-fit model parameters makes the {\tt LECRe} scenario unlikely to be a major process to account for the Sgr B2~X-ray emission in 2013.
If there is an underlying contribution from CR electrons, it has to be significantly lower than the current level.

\subsubsection{Self-consistent Cosmic Ray Proton Model {\tt LECRp}}

While CR electrons can be safely excluded as a major contributor to the 2013 Sgr~B2 X-ray emission, low energy CR proton/ion bombardment could be a major process if the Sgr~B2 X-ray emission obtained in 2013 has already reached the constant background level.
The {\tt LECRp} model has the same model parameters as the {\tt LECRe} model, with four free parameters $s$, $\Lambda$, $E_{min}$ and $N_{LECR}$.
As the path length cannot be constrained by the fit, we fix it to $\Lambda=5\times10^{24} \rm ~H-atoms~cm^{-2}$, a typical value for nonrelativistic particles propagating in massive molecular clouds of the GC environment \citep{Tatischeff2012}.
The right panel in Figure~4 shows that it results in an overall good fit with $\chi^2_{\rm \nu}=1.17$ for $d.o.f.=92$ (Table~2).
The best-fit Fe abundance is $Z/Z_{\sun}=2.5^{+1.5}_{-1.0}$, which is consistent with the GC metallicity.
The CR proton spectral index is $s=1.9^{+0.8}_{-0.7}$.  
This agrees with the CR proton spectral index derived with the {\tt LECRp} model for the arches region using the \xmm\ spectrum ($s=1.9^{+0.5}_{-0.6}$, \citealp{Tatischeff2012}) or the \nustar\ spectrum ($s=1.7\pm0.6$, \citealp{Krivonos2014}) 
CR iron bombardment was proposed by these authors to explain the X-ray emission from the Arches before a significant variability of the X-ray emission was detected \citep{Clavel2014}. 
As discussed in \citet{Tatischeff2012}, with such a relatively hard CR spectrum, the X-ray emission produced by CR protons depends weakly on the minimum ion energy $E_{min}$.
Therefore, we fix it to $E_{min}=10\rm~MeV~nucleon^{-1}$.
The total power required by CR protons in the cloud can be derived from the best-fit model normalization $N_{LECR}=(1.4\pm0.4) \times10^{-7}\rm~erg~cm^{-2}~s^{-1}$.
The power injected by primary protons of energies between $E_{min}=10\rm~MeV$ and $E_{max}=1\rm~GeV$ is $dW/dt=(1.0\pm0.3) \times10^{39}\rm~erg~s^{-1}$ for $D=7.9\rm~keV$.
However, the power injected into the cloud by the CR protons depends on the minimum energy $E_{min}$.
For $E_{min}=1\rm~MeV$ and $E_{min}=100\rm~MeV$, the best-fit model normalizations are $N_{LECR}=2.0^{+1.1}_{-0.5}\times10^{-7}\rm~erg~cm^{-2}~s^{-1}$ and $N_{LECR}=0.7^{+0.3}_{-0.1}\times10^{-7}\rm~erg~cm^{-2}~s^{-1}$, respectively.
The corresponding power injected by CR protons is in the range of $dW/dt=(0.4-2.3) \times10^{39}\rm~erg~s^{-1}$.
Therefore, the LECRp scenario could explain the current level Sgr~B2 X-ray emission and can only be excluded by further variability.

\subsection{Main Compact Cores Sgr~B2(M) and Sgr~B2(N)}

{\it Herschel} observations have shown that the local $250\mu \rm m$ optical depth at Sgr~B2(M) and Sgr~B2(N) ($\tau_{250\mu \rm m} \sim 1$) is higher than that of the surrounding region within 90\asec\ roughly by a factor of 2--5 \citep{Etxaluze2013}.
Specifically, the Sgr~B2(M) $250\mu \rm m$ optical depth is $\tau_{250\mu \rm m}\sim 0.9$.
\citet{Etxaluze2013} converted it to a local column density of $N_{H} \sim 1\times 10^{25}\rm~cm^{-2}$ based on the the conversion relationship $\tau_{250 \mu m}/N_{H}=8.8\times10^{-26}\rm~cm^{2}$ for optically thin clouds.
This conversion factor is increased to $5\times10^{-25}\rm~cm^{2}$ for the Sagittarius region \citep{Bernard2010}, thus reducing the Sgr~B2(M) column density to $N_{H} \sim1.8 \times 10^{24}\rm~cm^{-2}$.
We extracted a spectrum from a 20\asec\ radius region (Figure~2) centered on Sgr~B2(M) to measure the local optical depth independently using the X-ray data.    
The Sgr~B2(M) spectrum is fit with the {\tt MYTorus} model using the same model settings for the central 90\asec\ region.
The model yields a good fit with $\chi^2_{\rm \nu}=1.06$ for $d.o.f.=39$.
While all the other key parameters are consistent with those derived from the central 90\asec\ region, the intrinsic equatorial column density is $N_{H}(i)=(2.0^{+3.2_{stat}+0.5_{sys}}_{-1.0_{stat}-0.5_{sys}}) \times 10^{24}\rm~cm^{-2}$, twice that of the whole 90\asec\ region but compatible within error bars.
The {\tt LECRP} model can also fit well to the Sgr~B2(M) spectrum ($\chi^2_{\rm \nu}=1.02$ for $d.o.f.=39$), giving consistent $N_{H}(i)$ value.
It agrees well with the {\it Herschel} measurements of local column densities.
The corresponding radial optical depth in the direction of Sgr~B2(M) is $\tau_{T}=1.3^{+2.5}_{-1.0}$.
The \nustar\ observations confirm the {\it Herschel} measurements that the Sgr~B2(M) region is not optically thin, though with large error bars.
The 10--40~keV flux of Sgr~B2(M) and Sgr~B2(N) falling into the central 90\asec\ region is $F_{10-40~keV}=(6.1\pm0.2)\times10^{-13}\rm~erg~cm^{-2}~s^{-1}$, contributing to about one third of the total flux of the central 90\asec.
The surrounding regions within the 90\asec\ radius region contribute to the remaining two thirds of the total flux.
The second core, Sgr~B2(N), does not have sufficient statistics to perform a similar analysis.

\subsection{The new feature G0.66$-$0.13}
 
Since neither focal plane modules are severely contaminated by stray light, we were able to use the combined FPMA and FPMB \nustar\ data for G0.66$-$0.13.
Furthermore, the G0.66$-$0.13 region does not contain sources registered in the \chandra\ point source catalogue \citep{Muno2009}. 
The background-subtracted \nustar\ spectrum for G0.66$-$0.13 does not show any significant line features.
To compare with the significant Fe K$\alpha$ lines detected by \xmm\ in 2012, we fit the \nustar\ 2013 spectrum together with the \xmm\ 2012 spectrum.
Since the sharply decreasing Fe K$\alpha$ emission cannot be explained by the LECRp scenario, we examine the G0.66$-$0.13 spectrum only with the {\tt MYTorus} model.  
Because of different background subtraction methods, the thermal component in the \xmm\ data represents the GC diffuse emission while the thermal component in the \nustar\ data represents undetected point source and residual diffuse emission.
Therefore we use the model {\tt wabs*(apec+MYTS+MYTL)} for the joint fitting, with all the model parameters linked except for $kT$ and the model normalizations of {\tt apec}, {\tt MYTS} and {\tt MYTL}.  
The model gives a good fit ($\chi^2_{\rm \nu}$=1.18 for $d.o.f.=55$), with a best-fit photon index of $\Gamma=1.4\pm0.5$, a foreground column density of $N_{H}(f)=(8.2^{+4.3}_{-4.5})\times10^{22}\rm~cm^{-2}$ and an intrinsic equatorial column density of $N_{\rm H}(i)=\rm 3.0^{+3.8}_{-1.9} \times 10^{23}\rm~cm^{-2}$ (Figure~5).
The intrinsic density $N_{H}(i)$ is lower than that of the central 90\asec\ radius region, indicating that G0.66$-$0.13 is optically thin.
However, in case that G0.66$-$0.13 is partially illuminated, the measured $N_{H}(i)$, which is the illuminated column density, would be smaller than the cloud intrinsic column density.  
The 10--40~keV flux for the \nustar\ 2013 data is $F_{10-40\rm~keV}=(9.0 \pm 1.2) \times 10^{-13} \rm~erg~cm^{-2}~s^{-1}$ in 10--40~keV.
The observed 6.4~keV flux in 2012 measured by \xmm\ is $F_{6.4\rm~keV}=(1.5\pm0.2) \times 10^{-5}\rm~ph~cm^{-2}~s^{-1}$,
while the 2013 \nustar\ spectrum does not require a 6.4~keV line, with 6.4~keV flux upper limit of $5\times10^{-6}\rm~ph~cm^{-2}~s^{-1}$ (90\% confidence level).
The 8--12~keV non-thermal continuum emission measured by \nustar\ in 2013 dropped to $\sim 50\%$ of that measured by \xmm\ in 2012.
Both the fluorescence emission and the hard X-ray emission show fast variability (or short life time), which could be due to a partial illumination of G0.66$-$0.13.

\begin{figure*}
\centeringÊ 
\label{fig:spectra_SgrB2}
\begin{tabular}{cc}
\includegraphics[angle=0, width=1.0\textwidth]{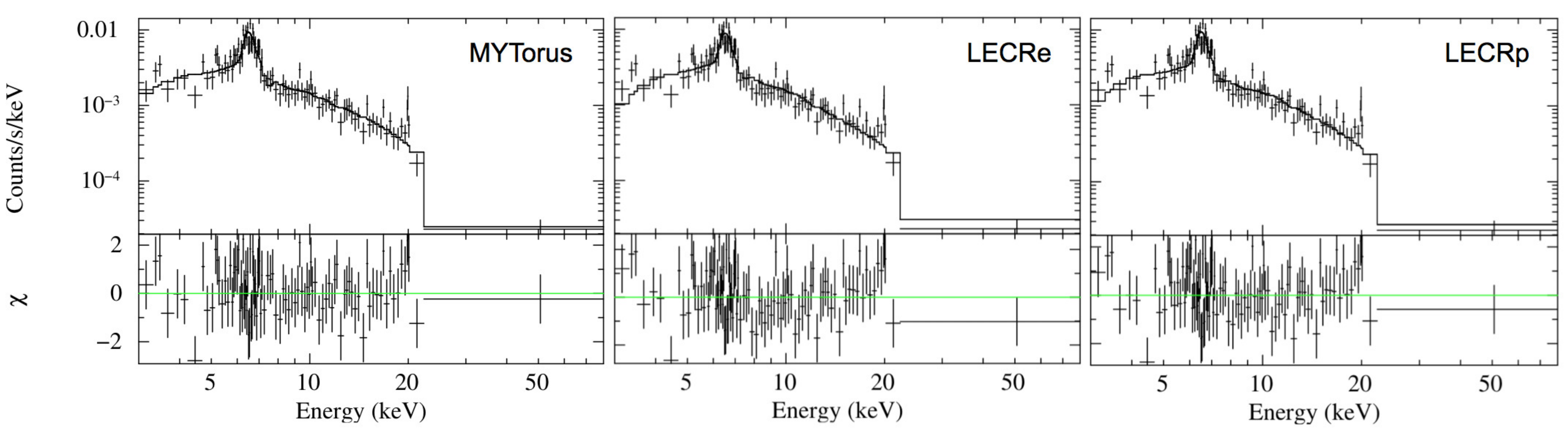}
\end{tabular}
\caption{The 3--79~keV X-ray spectrum of the central 90\asec\ of \sgrb\ measured with \nustar\ FPMA fitted with three different models. The crosses show the data points with $1 \sigma$ error bars. The solid lines are the best-fit models. The lower part of the plots shows the deviation from the model in units of standard deviation. The spectrum is fitted with the self-consistent XRN model {\tt MYTorus} (left panel), the {\tt LECRe} model (middle panel) and the {\tt LECRp} model (right panel). While all the models can fit the spectrum well overall, the {\tt LECRe} model results in a poorer fitting above 10~keV. The {LECRe} scenario is ruled out based on unphysical best-fit model parameters.}
\end{figure*}

\begin{figure*}
\centeringÊ 
\label{fig:spectra_G066}
\begin{tabular}{cc}
\includegraphics[angle=270, width=0.4\textwidth]{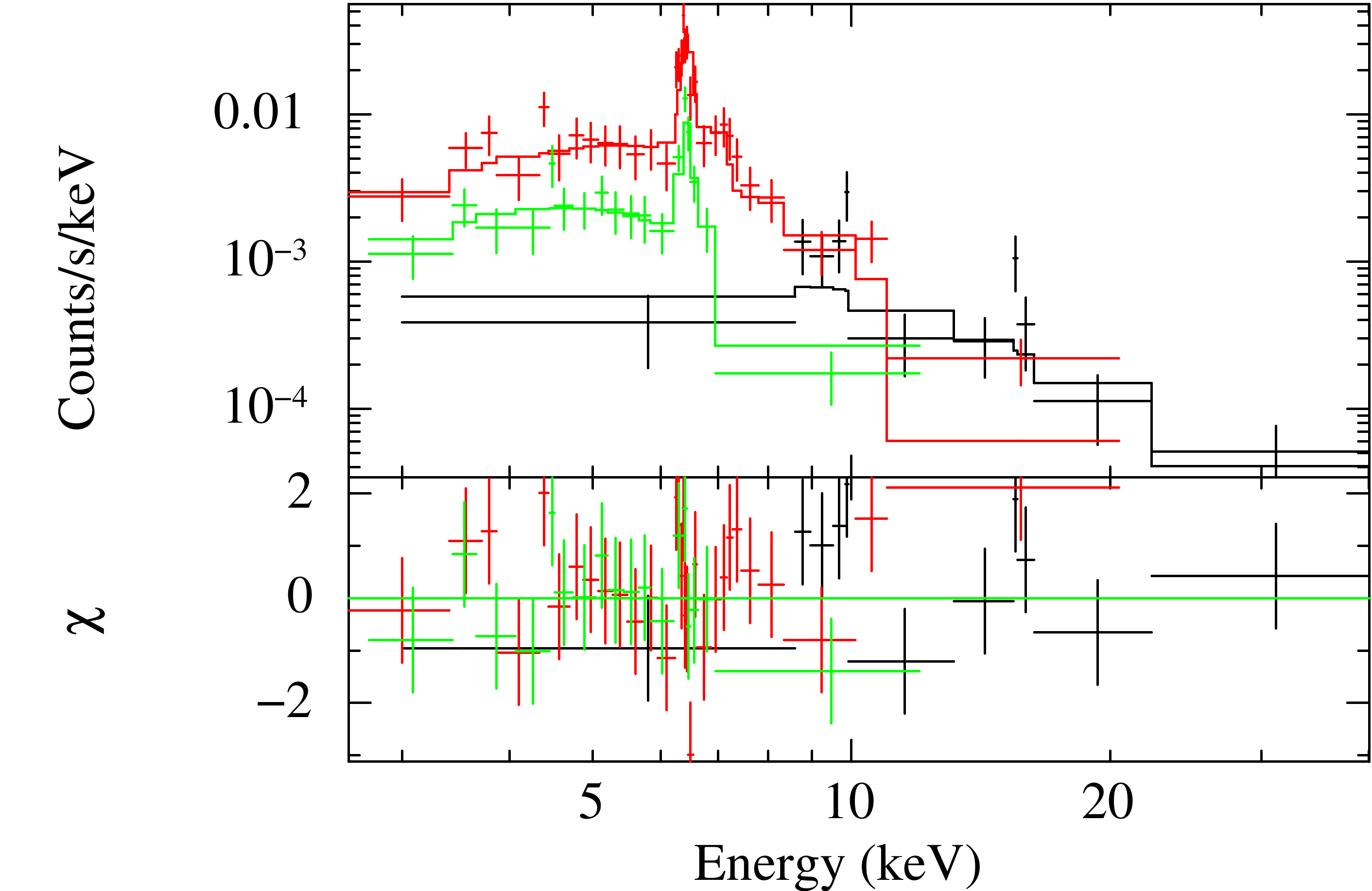} 
\end{tabular}
\caption{2013 \nustar\ (black) and 2012 \xmm\ (red for PN and green for MOS) spectra of G0.66$-$0.13 jointly fit to the XRN model {\tt wabs*(apec+MYTS+MYTL)}. The crosses show the data points with $1 \sigma$ error bars, and the solid lines show the best model. The lower panel shows the deviation from the model in units of standard deviation. The 6.4~keV line was significantly detected by \xmm\ in 2012, but not detected by \nustar\ in 2013. It suggests a fast decay of fluorescence emission within one year.}
\end{figure*}

\begin{deluxetable*}{lccccc}                                                                                                               
\tablecaption{Spectral analysis of central 90\asec\ of \sgrb\ with {\it NuSTAR} data.}
\tablecolumns{5}                                                                                                                    
\tablehead{ \colhead{Parameter}      & \colhead{Unit}                 &  \colhead{Power-law~$\rm Model^{a}$}   & \colhead{$\tt MyTorus\rm~Model^{b}$}   & \colhead{\tt LECRe~$\rm Model^{c}$}   & \colhead{\tt LECRp~$\rm Model^{d}$}}
\startdata
$N_{\rm H}(f)$                                 & 10$^{23}$~cm$^{-2}$       & $1.1 \pm 0.4$                                            & $1.2 \pm 0.1$                                        &  $1.3 \pm 0.5$                			           &  $1.2\pm0.6$  \\
$N_{\rm H}(i)$                                 & 10$^{23}$~cm$^{-2}$       & $5.0\pm1.3$                                              & $10.1\pm4.1$                                        &  $5.5\pm1.3$                                          &  $6.5^{+3.9}_{-2.6}$ \\
$Z/Z_{\sun}$                                   &                                           &   \nodata                                                    & $1$(fixed)                                              &  $4.0^{+2.0}_{-0.6}$                               &  $2.5^{+1.5}_{-1.0}$ \\
& & & & & \\  
$kT$                      			        & keV                                    & $2.1^{+1.1}_{-0.9}$                                   & $3.5^{+2.5}_{-1.4}$                               & $3.4^{+2.3}_{-1.2}$                                & $2.6^{+1.4}_{-0.7}$     \\
$F_{\rm apec}(2-10\rm~keV)$     & $\rm erg~cm^{-2}~s^{-1}$    & $6^{+3}_{-4} \times10^{-13}$                    & $6^{+2}_{-3}\times 10^{-13}$                 & $7^{+4}_{-3} \times10^{-13}$                 &                                     \\
& & & & & \\
$F_{6.4\rm keV}$		   & 10$^{-6}$ ph~cm$^{-2}$~s$^{-1}$     & $8.3\pm2.1$	                       			           & \nodata                                                  & \nodata  									& \nodata  \\
$\rm EW_{6.4\rm~keV}$                & keV                                    & $1.2^{+0.7}_{-0.3}$						     & \nodata						                 & \nodata   								      & \nodata   \\
& & & & & \\
$\Gamma_{p.l.}$                            &                                           & $1.9\pm0.5$  		       		                 & \nodata                                                  & \nodata  								      & \nodata   \\
$F_{p.l.}(10-40\rm~keV)$      & $\rm erg~cm^{-2}~s^{-1}$         &  $(1.9\pm{0.2}) \times10^{-12}$	   	           & \nodata                                	                 & \nodata        							      & \nodata    \\
& & & & & \\
$\Gamma_{MT}$                          &							     & \nodata									     & $2.2\pm0.4$          					     & \nodata                                                   & \nodata    \\
$N_{MT}$            & $10^{-2}\rm~ph~cm^{-2}~s^{-1}$                & \nodata								           & $1.8^{+0.7}_{-1.1}$					     & \nodata                                                   & \nodata   \\
& & & & & \\
$\Lambda$                     & $\rm H-atoms~cm^{-2}$                   &   \nodata	                                                     & \nodata                                                 & $5\times10^{24}$ (fixed) 	                 & $5\times10^{24}$ (fixed) \\								      
$s$							        &   		                                  &    \nodata				                  		     & \nodata					                       & $2.8^{+0.4}_{-0.5}$                               &  $1.9^{+0.8}_{-0.7}$ \\
$E_{min}$                                 & keV                                        &     \nodata                                                     & \nodata                                                 & $3^{+47}_{-...}$                                      &  $10^{4}$ (fixed) \\
$N_{LECR}$    & $\rm 10^{-7}\rm~erg~cm^{-2}~s^{-1}$            &   \nodata				                  		     & \nodata					                       & $2.8\pm0.2$                                          & $1.4\pm0.4$ \\
& & & & & \\  
$\chi^2_{\rm \nu}$ (d.o.f.)        &                                               & 0.97 (91)                                                      & 1.09 (92)								     & 1.18 (92)                                               & 1.17 (92) \\
\enddata
\tablecomments{
The goodness of fit is evaluated by $\chi^2_{\rm \nu}$ and the number of degrees of freedom given in parentheses. 
The errors are 90\% confidence.\\
$^{a}$~Power-law Model: {\tt wabs*(apec+wabs*pow+gauss+gauss)}.  \\
$^{\ }$~Power-law model is characterized by foreground interstellar absorption column density $N_{\rm H}(f)$,\\ 
$^{\ }$~temperature $kT$ and $2-10$~keV flux $F_{apec}(2-10\rm~keV)$ of {\tt apec}, intrinsic absorption column density \\  
$^{\ }$~$N_{\rm H}(i)$, photon index $\Gamma_{p.l.}$ and $10-40$~keV observed flux $F_{p.l.}(10-40\rm~keV)$ of power-law, equivalent \\ 
$^{\ }$~width $EW$ and observed flux $F_{6.4\rm keV}$ of the 6.4~keV iron line.\\
$^{b}$~{\tt MYTorus} model: {\tt wabs*(apec+MYTorus)}. \\
$^{\ }$~The {\tt MYTorus} model is characterized by $N_{\rm H}(f)$, $kT$ and $F_{\rm apec}(2-10\rm~keV)$ of {\tt apec}, the intrinsic \\
$^{\ }$~absorption column density $N_{\rm H}(i)$, photon index $\Gamma_{MT}$ of the illuminating power-law spectrum \\
$^{\ }$~and the power-law normalization $N_{MT}$, while the inclination angle is fixed to $\theta_{obs}=0^{\circ}$. \\
$^{c}$~{\tt LECRe} model:  {\tt wabs*(apec+wabs*LECRe)}. \\
$^{\ }$~The {\tt LECRe} model is characterized by $N_{\rm H}(f)$,  $N_{\rm H}(i)$, $kT$ and $F_{apec}(2-10\rm~keV)$ of {\tt apec}, the power-law \\
$^{\ }$~index $s$ of the accelerated particle source spectrum varying from 1.5 to 5,  the path length $\lambda$ of the \\
$^{\ }$~cosmic rays in the non-thermal X-ray production region varying from $10^{21}$ to $10^{26}\rm~H-atoms~cm^{-2}$,  \\
$^{\ }$~minimum energy $\rm E_{min}$ and the model normalization $\rm N_{LECR}$ for {\tt LECRe}. $\rm N_{LECR}$ provides the power  \\
$^{\ }$~injected in the interaction region by primary cosmic-ray electrons or protons of kinetic energies  between \\
$^{\ }$~$E_{min}$ and $E_{max} = 1 \rm~GeV$. \\
$^{d}$~{\tt LECRp} model: {\tt wabs*(apec+wabs*LECRp)}. \\
$^{\ }$~The {\tt LECRp} model is characterized by $N_{\rm H}(f)$,  $N_{\rm H}(i)$, $kT$ and $F_{apec}(2-10\rm~keV)$ of {\tt apec},  $s$ ranging \\
$^{\ }$~from 1 to 5,  $\lambda$ ranging from $10^{21}$ to $10^{26}\rm~H-atoms~cm^{-2}$,  $E_{min}$ and the model $N_{LECR}$ for {\tt LECRp}. 
}
\label{tab:specfit}
\end{deluxetable*}


\section{Discussion}

\subsection{X-ray Reflection of Past Sgr~A$^\star$ Outbursts}    

The broadband spectrum obtained by \nustar\ supports the XRN scenario where \sgrb\ reflects incoming X-rays from an illuminating source, which is most likely a past X-ray outburst from Sgr~A$^\star$.
The illuminating source photon index measured with the broadband \nustar\ spectrum in 2013 using the {\tt MyTorus} model is constrained to $\Gamma=2.2\pm0.4$, compatible with the illuminating source photon index of $\Gamma=2.0\pm0.2$ derived with combined \xmm\ and \integral\ spectra obtained in 2003-2004 when Sgr~B2 was much more brighter \citep{Terrier2010}.
Such a past Sgr~A$^\star$ outburst spectrum is consistent with the current bright Sgr~A$^\star$ X-ray flares spectra with $\Gamma \approx 2-3$ (e.g. \citealp{Baganoff2001, Nowak2012, Barriere2014}), although the mechanisms of both the past giant X-ray outbursts and the current relatively mild X-ray flares remain unclear.
The outburst spectrum is also compatible with, though at the extreme end of, the average AGN photon index of $1\le \Gamma \le 2$ with cut-off energies between 30~keV to 300~keV \citep{Molina2013}, which is also expected for the
Sgr A$^\star$ spectrum if it became active.

The \sgrb\ cloud as a whole has been discussed as an optically thin cloud (e.g. \citealt{Koyama1996, Sunyaev1998, Revnivtsev2004, Terrier2010}).
In 2013, \nustar\ found the hard X-ray emission is concentrated in the central 90\asec\ radius region of \sgrb, but it is unknown whether the hard X-ray emission in 2003-2009 was similarly concentrated within 90\asec\ or more extended due to the mild angular resolution of \integral.
While in the direction of the compact core Sgr~B2(M) the optical depth is as high as $\tau_{T} \gtrsim 1$, the majority of the central 90\asec\ radius region is optically thin with $\tau_{T} < 1$.
Besides, the 10$-$40~keV image shows that the X-ray continuum emission traces the local column density, also suggesting the majority of the central 90\asec\ of \sgrb\ is optically thin.
Therefore, although there are complicated substructures with higher local column densities, the central 90\asec\ can be considered as optically thin for a simplified calculation of Sgr~A$^\star$ outburst luminosity.

For an optically thin cloud, the luminosity of the primary source can be derived via two independent analytical methods: the Fe K$\alpha$ fluorescence process using the 6.4~keV line flux we measured, and the Compton scattering process with the measured hard X-ray continuum.
In most previous works, the primary source luminosity has been calculated via the fluorescence process, due to good measurements of the 6.4~keV line but poorly constrained continuum emission.
The 6.4 keV line emission flux for continuum radiation is expressed by \citet{Sunyaev1998} as

\begin{equation}
\begin{split}
F_{6.4}=\phi \frac{\Omega}{4\pi D^2} Z_{Fe} \tau_{T} I(8) ~~\rm ph~cm^{-2}~s^{-1}
\end{split} 
\end{equation}
where $\Omega$ is the cloud solid angle subtended to the illuminating source, $D$ is the distance to the observer, 
$Z_{Fe}$ is the iron abundance relative to solar, $\tau_{T}$ is optical depth due to Thomson scattering, and $I(8)$ is the illuminating source flux at 8 keV. 
$\phi$ is a factor of the order of unity, which depends on the source spectrum. 
\citet{Sunyaev1998} assumed the source spectrum to be bremsstrahlung with a temperature from 5 to 150 keV, corresponding to $\phi \sim 1.0-1.3$.
Now, with a better knowledge of the primary source spectrum, we know it can be well described with a power-law with $\Gamma \approx 2$.
$\Gamma$ ranges from 1.8 to 2.6, as measured with the {\tt MYTorus} model, corresponding to $\phi \sim 1.1-1.3$.
In the calculation hereafter we will adopt $\Gamma \approx 2$ and its corresponding $\phi=1.18$.

Using the \citet{Sunyaev1998} approach, we express $I(8)$ with illuminating source luminosity at 8 keV in the 8 keV wide energy band as $L_{8}=I(8) \times 8 \times 8 \times 1.6 \times 10^{-9}~\rm erg~s^{-1}$.
For the central region of \sgrb\ we adopted $r=90\asec$ assuming it was entirely illuminated, so the luminosity of the source required to produce the observed 6.4~keV line is:

\begin{equation}
L_{8}=3 \times 10^{39} \left(\frac{F_{6.4}}{10^{-4}}\right) \left(\frac{0.1}{\tau_{T}}\right) \left(\frac{d}{100\rm~pc}\right)^{2} (Z_{Fe})^{-1}~~\rm erg~s^{-1}
\end{equation}
where $d$ is the distance between the molecular cloud and the illuminating source.
For a power-law spectrum with $\Gamma \approx 2$, $L_{8}$ corresponds to 30\% of the source luminosity at $3-79$~keV.  
While the 6.4~keV line flux has been decreasing over the past decade, the maximum observed Fe K$\alpha$ flux should be used to calculate the peak primary source luminosity.
As the XRN model we use does not measure the iron abundance of \sgrb, we used the most recently measured value $Z_{Fe}=1.3\pm0.1$ \citep{Terrier2010}.  
With the maximum central 90\asec\ Sgr~B2 flux $F_{6.4}=(4.3\pm0.6) \times 10^{-5} \rm~ph~cm^{-2}~s^{-1}$ obtained in 2001 and averaged optical depth $\tau_{T}=0.67\pm0.27$ measured with by \nustar, the resultant primary source luminosity is $L_{8} =(1.5\pm0.7) \times10^{38}(d/100~\rm pc)^2~\rm erg~s^{-1}$, corresponding to a 3--79 keV luminosity of $L_{3-79\rm~keV}=(5.0\pm2.3) \times 10^{38}(d/100~\rm pc)^2~\rm erg~s^{-1}$ or 1--100~keV luminosity of $L_{1-100\rm~keV}=(0.7\pm0.3) \times 10^{39} (d/100~\rm pc)^2\rm~erg~s^{-1}$.
This is comparable to the $L_{1-100\rm~keV}\sim1.1\times 10^{39} (d/100~\rm pc)^2\rm~erg~s^{-1}$ derived from \xmm\ and \integral\ data obtained in 2003-2004 \citep{Terrier2010}.
The uncertainty of the primary source luminosity mainly comes from the measurement uncertainty of $\tau_{T}$ and the distance between \sgrb\ and the primary source, $d$.
The solid angle $\Omega$ could also be lower in the case of partial illumination, which would result in an underestimation of the primary source luminosity.

Since this was the first time the \sgrb\ 10--40~keV continuum emission has been resolved, we can also derive the primary source continuum luminosity from the Compton scattered continuum.
As discussed in \citet{Capelli2012}, the observed scattered continuum at an angle $\theta$ with respect to the incident radiation direction is:
\begin{equation}
S_(E)=\frac{\Omega}{D^2} \frac{3\tau_{T}}{8\pi} \frac{(1+\cos^{2}\theta)}{2} I(E)~\rm ph~cm^{-2}~s^{-1}~keV^{-1}
\end{equation}
where $I(E)$ is the photon flux from the source in units of $\rm ph~s^{-1}~keV^{-1}$.
This relationship is based on Thomson scattering, which is a good approximation for incident photons with energies of a few keV.
However, when the photon energy is higher, electron quantum mechanical effects become important.
As a result, we need to consider the Compton scattering process.
Starting with the Compton scattering cross section formula, we derived a correction factor from Thomson scattering to Compton scattering as $f_{TC}=(1+\frac{E}{511\rm~keV}(1-\cos \theta))^{-4}$ assuming a power-law spectrum with $\Gamma \approx 2$.
The correction holds as long as $\hbar \omega \ll mc^{2}$ for the electrons, which is true for the \nustar\ energy band of 3 to 79 keV.

Thus, for any X-ray photon energy, the Compton scattered continuum can be expressed as:

\begin{eqnarray}
S(E)&=&\frac{\Omega}{D^2} \frac{3\tau_{T}}{8\pi} \frac{(1+\cos^{2}\theta)}{2} \left(1+\frac{E}{511\rm~keV}(1-\cos \theta)\right)^{-4} \nonumber \\ 
&&\times I(E) \rm~ph~cm^{-2}~s^{-1}~keV^{-1}
\end{eqnarray}
The observed 10--40~keV scattered continuum flux is $F_{10-40}=\int_{10\rm~keV}^{40\rm~keV} \! S(E) E\, \mathrm{d}E ~\rm erg~cm^{-2}~s^{-1}$. 
Thus, integrating both sides of equation (4) gives us the relationship between $F_{10-40}$ and $L_{8}$.
We keep the first order correction term of the integrated form and neglect higher order terms.
As a result, the source luminosity in the 8~keV wide band can be expressed with observed 10--40~keV Compton scattered continuum measured at angle $\theta$ as: 

\begin{eqnarray}
L_{8}&=&4\times 10^{39} \left(\frac{F_{10-40}}{10^{-11}\rm erg~cm^{-2}~s^{-1}}\right)  \left(\frac{0.1}{\tau_{T}}\right) \left(\frac{d}{100\rm~pc}\right)^2 \nonumber \\ 
&&\times \frac{1}{1+ \cos^{2} \theta} \times (0.72+0.12(1-\cos \theta))~\rm erg~s^{-1}
\end{eqnarray}
We use the maximum value of the hard X-ray continuum flux in 2012 from the lightcurve obtained by \xmm.   
In Section 5 we estimated the 10--40~keV flux of the central 90\asec\ in 2001.
The resultant source luminosity is $L_{8}=(3.6\pm1.8) \times 10^{38}~f(\theta)~(d/100\rm~pc)^2~\rm erg~s^{-1}$, where $f(\theta)=\frac{1}{1+\cos^2 \theta} (0.72+0.12(1-\cos \theta)) \sim$ 0.36--0.84 corresponding to $\theta \sim 0^{\circ}$--$180^{\circ}$. 
As the scattering angle range cannot be derived from the XRN model fitting in our study, we adopt $\cos \theta \sim 0.8$ based on VLBI observation results \citep{Reid2009} to derive a nominal value for $f(\theta)$.
The error of $f(\theta)$ is calculated corresponding the full allowed range of the scattering angle $\theta=0^{\circ}$--$180^{\circ}$, resulting in $f(\theta)=0.46^{+0.38}_{-0.10}$.
Thus, $L_{8}=(1.7^{+1.6}_{-0.9}) \times 10^{38}~(d/100\rm~pc)^2~\rm erg~s^{-1}$, corresponding to a 3--79 keV luminosity of $L_{3-79\rm~keV}=(5.6^{+5.2}_{-2.9}) \times 10^{38}(d/100~\rm pc)^2~\rm erg~s^{-1}$ or a 1--100~keV luminosity of $L_{1-100\rm~keV}=(0.8^{+0.7}_{-0.4}) \times 10^{39} (d/100~\rm pc)^2\rm~erg~s^{-1}$.

This result is consistent with $L_{8}$ calculated via the photoelectric absorption process. 
Indeed, comparing equation (2) and (5), we have

\begin{equation} 
\frac{L_{8}(abs)}{L_{8}(scat)}= \frac{3 \left(\frac{F_{10-40}}{10^{-11}\rm erg~cm^{-2}~s^{-1}}\right) (Z_{Fe})^{-1}} {4 \left(\frac{F_{6.4}}{10^{-4} \rm ph~cm^{-2}~s^{-1}}\right) f(\theta)}=0.8^{+0.4}_{-0.6}
\end{equation} 
which is a factor of the order of unity. 
The primary source luminosity independently calculated via the Compton scattering is remarkably consistent with that calculated via photoelectric absorption.
The consistency strongly supports the X-ray reflection scenario.
With Equation (4), the primary source luminosity in any energy band can be directly derived given the measurement of the molecular cloud flux in the same energy band.
This continuum-flux-based method has also been applied to other GC molecular clouds to test the XRN model and constrain the Sgr $A^{\star}$ outburst history (Mori et al. 2015).

The {\tt MYTorus} model can also give the illuminating source spectrum, although it assumes fixed solar abundance for iron and that the observation angle cannot be constrained with the data.
By rescaling the incoming X-ray flux by the solid angle ratio of the torus ($\Omega/4\pi=0.5$) and Sgr~B2 ($\Omega/4\pi=2.89\times10^{-4}$), we derive the primary source luminosity to be $L_{8}=(1.4^{+0.5}_{-0.1})\times 10^{38}~(d/100\rm~pc)^2~\rm erg~s^{-1}$ (see the Appendix for details),
As the 2013 X-ray emission from \sgrb\ is 30\% that of its maximum flux (Section 5), the primary source luminosity with the peak Sgr~B2 flux is $L_{8}=(4.6^{+1.6}_{-3.6})\times 10^{38}~(d/100\rm~pc)^2~\rm erg~s^{-1}$, or $L_{3-79\rm~keV}=(1.5^{+0.5}_{-1.1}) \times 10^{39}(d/100~\rm pc)^2~\rm erg~s^{-1}$ with the 2001 \sgrb\ X-ray flux.
With large error bars, the primary source luminosity derived by the {\tt MYTorus} model overlaps with that derived with the analytical calculations, though the best-fit luminosity value is higher.
For the other GC molecular clouds, the {\tt MYTorus} model also gives a primary source luminosity higher than that derived from the analytical calculation roughly by a factor of 2 (Mori~et al. 2015).

We note that the analytical calculations above are valid for an optically thin ($\tau_{T} \ll 1$) cloud where the fluorescence photons are not scattered and the continuum emission photons are scattered once inside the cloud.
Sgr~B2 has very complicated density profile, including three main compact cores (Sgr~B2(N), Sgr~B2(M) and Sgr~B2(S) from north to south) and numerous clumpy regions.
Overall its density decreases as radius becomes larger. 
Without a model fully considering the density distribution within \sgrb, the measured $\tau_{T}$ for a specific region is an averaged value.
For the 90\asec\ radius region we adopt, and larger regions previous works adopt (e.g. 4.5\amin\ radius region in \citealp{Terrier2010}), the averaged optical depth for such large regions is $\tau_{T} \ll 1$, thus have been treated as optically thin as a whole.
However, for the compact cores at the center of Sgr~B2, which have higher local density, the local optical depths are higher than 1 as indicated by {\it Herschel} and \nustar\ measurements, though with large error bars.
For these regions, multi-scattering could take place during the decay stage.
The X-ray emission life time of optically thick material is longer than its light crossing time ($t=\tau_{T} r/c$ when $\tau \gg 1$) due to decay caused by multi-scattering.  
The reflected X-ray emission follows an exponential decay for optically thick materials and a parabolic decrease for optically thin materials.
Therefore, the optically thicker material flux decreases much slower at late stages.  
If the compact core regions are confirmed to be optically thick, X-ray emission from the central compact cores Sgr~B2(M) and Sgr~B2(N) could still be bright enough to be detected when the surrounding optically-thin regions completely fades.
However, if the optical depth in the direction of the compact cores can be more tightly constrained to $\tau_{T}=1$, the multi-scattering effect would be negligible. 
Future observation of \sgrb\ can test this prediction and better constrain the local optical depth in the direction of the compact cores.

\subsection{Cosmic Ray Electron Bombardment}

LECRe cannot be the major contribution to the bright Fe K$\alpha$ line emission from \sgrb, as both the required iron abundance and the electron power are too high to be physical (e.g. \citealt{Revnivtsev2004, Terrier2010}).
After a decade of decreasing in the Fe K$\alpha$ line flux, we tested whether the the constraints on the LECRe parameters by the remaining emission level in 2013 are still excluding this scenario.
Generating the bremsstrahlung emission with the observed slope of $\Gamma=1.9$ in 3--79~keV requires a relatively soft electron spectra with $s \sim 2.8$ and a minimum CR electron energy far below 100 keV as we measured, which is illustrated in \citet{Tatischeff2012}. 
For such a soft CR spectrum, the neutral Fe K$\alpha$ line is predicted to be relatively weak, with $\rm EW<0.4 \times (Z/Z_{\sun})$~keV (Fig.~3 in \citealt{Tatischeff2012}). 
Therefore, to fit the observed Fe K$\alpha$ line EW of $\sim1.3$~keV as measured by \nustar, the required Fe abundance is found to be $Z_{Fe}=4.0^{+2.0}_{-0.6}$, still significantly exceeding the current measurement of GC metallicity of solar to twice solar.
This is due to the relatively inefficient production of 6.4~keV photons by LECRe interactions, for which the iron fluorescence yield $R_{6.4\rm~keV}=L_{X}(6.4\rm~keV)/(\mathrm{d}W_{e}/\mathrm{d}t)$ is always lower than $1 \times 10^{-6}(Z/Z_{\sun})$ for $s>2.3$ and $E_{min}<100~\rm keV$ \citep{Tatischeff2012}. 
Besides the unphysical Fe abundance required by the Fe K$\alpha$ emission, the {\tt LECRe} model also results in an overall poorer fitting to the broadband 3--79~keV spectrum compared to the XRN model.
Especially, the hard X-ray part of the spectrum has more residuals for the {\tt LECRe} model, thus providing a more exclusive constraint which does not depend on local metallicity.

Another difficulty in explaining all the 6.4 keV emission with LECRe is the lack of an obvious particle acceleration site. 
\citet{Yusef2002} suggested that the interaction of non-thermal radio filaments and the cloud could locally accelerate CR electrons.
Yet until now no such filaments have been observed around \sgrb.   
Even if there exists a local CR acceleration site, the non-thermal electrons with energies below 100~keV are not able to escape the acceleration region and penetrate the whole dense cloud \citep{Yusef2007}.
Therefore, we conclude that LECRe scenario cannot be a major contributor to the remaining X-ray emission from \sgrb\ in 2013.

\subsection{Cosmic Ray Ion Bombardment}

Although it is clear that the Sgr~B2 Fe K$\alpha$ emission decreases since 2001, 
The same level of the Fe K$\alpha$ emission in 2012 and 2013, and the lacking of measurements between 2005 and 2011 do not allow us to determine whether the Fe K$\alpha$ emission was still decreasing in 2013 or it has reached the constant background emission level. 
Future observations of Sgr~B2 can distinguish between the two cases.
\citet{Dogiel2009} estimated that CR protons could contribute to about 15\% of the observed maximum Fe K$\alpha$ flux obtained around 2000, although this prediction is highly model dependent. 
In 2012 and 2013, the Sgr~B2 Fe K$\alpha$ emission reached 20\% of that measured in 2001 (Section 4), therefore the 2012 and 2013 Fe K$\alpha$ emission could be mainly due to CR proton bombardment. 

The 2013 \nustar\ Sgr~B2 spectrum can be fit well with the self-consistent LECRp model.
The required metallicity of $Z/Z_{\sun}=2.5^{+1.5}_{-1.0}$ overlaps with the GC environment metallicity of one solar to twice solar.
Assuming all the X-ray emission from the central 90\asec\ Sgr~B2 in 2013 is due to CR proton bombardment, the required total power of the CR proton of energies between $E_{min}=10\rm~MeV$ and $E_{max}=1\rm~GeV$ in the cloud region is $dW/dt=(1.0\pm0.3) \times 10^{39}\rm~erg~s^{-1}$.
Taking the uncertainty of $E_{min}$ into account, the required CR proton energy ranges in $dW/dt=(0.4-2.3) \times 10^{39}\rm~erg~s^{-1}$ (Section 5.1.4).
According to \citet{Tatischeff2012}, there is an additional 40\% power comes from $\alpha$-particles with $C_{\alpha}/C_{p}=0.1$, the final required total kinetic CR ion power is $(0.6-3.2) \times 10^{39}\rm~erg~s^{-1}$, which is roughly 10\% of the steady state mechanical power supplied by supernovae in the inner $\sim 200$~pc  of the Galaxy \citet{Tatischeff2012}.
The required power injection is quite significant and and probably difficult to fully be accounted with typical sources.

The power deposited into the cloud is lower than the incident CR ion power, because CR ions with energies lower than $E_{min}$ cannot penetrate the cloud and that with a path length of $\Lambda=5 \times 10^{24}\rm~cm^{-2}$ those ions with energies higher than 180~MeV can escape from the cloud without depositing energy in it \citep{Tatischeff2012}.
Therefore, the power deposited by CR ions into central Sgr~B2 is $\dot{W}_{d} \sim 8 \times 10^{38}\rm~erg~s^{-1}$.
With the central 90\asec\ of Sgr~B2 mass of $M \sim (0.5-2) \times 10^{6} M_{\sun}$ estimated based on the simplified Sgr~B2 density profile (Section 1) and a total mass of $M=6\times10^{6} M_{\sun}$ \citep{Lis1990}, 
the corresponding ionization rate can be estimated to be $\zeta_{H} \sim (6-10) \times 10^{-15}\rm~H^{-1}~s^{-1}$ using Equation (11) in \citet{Tatischeff2012}.
This CR ionization for the dense materials in Sgr~B2 is comparable to the GC CR ionization rate of $\zeta_{H} \sim (1-3)\times10^{-15}\rm~s^{-1}$, which is found to be uniform throughout the GC on scales of 200~pc (e.g. \citealp{Goto2011}).

In the following we compare the application of LECRp model on Sgr~B2 and the Arches region and discuss possible LECR ion sources. 
The X-ray emission from the Arches region was interpreted as CR ion bombardment before a significant decrease in Fe K$\alpha$ was revealed \citep{Clavel2014}.
In the LECRp scenario, the CR spectral index is constrained by the broadband \nustar\ Arches region spectrum to $s=1.65^{+0.59}_{-0.55}$ \citep{Krivonos2014}, which agrees with that derived from the Sgr~B2 specrtum ($s=1.9^{+0.8}_{-0.7}$).
The GC LECR ion source is suggested to be Galactic supernovae or star accretion onto Sgr~A$^{\star}$ (\citealp{Dogiel2013} and the references therein).
The arches spectrum requires an ionization rate ($\zeta_{H} \sim 10^{-13}\rm~H^{-1}~s^{-1}$, \citealt{Tatischeff2012}) higher than that of Sgr~B2 by a factor of $\sim10-20$, significantly exceeding the CR ionization rate in the GC environment.
Recently, \citet{Clavel2014} found a 30\% flux drop from the Arches cluster in 2012, which indicates that a large fraction of the non-thermal emission of the Arches is likely due to the past activity of Sgr A$^{\star}$.
This could explain the high ionization required by the Arches spectrum, assuming all the X-ray emission is due to LECRp.
If the CR ionization rate of the Arches cluster is comparable to the GC CR ionization rate, LECRp bombardment should contributes to only a few percent of the total X-ray emission.
However, there could be local LECR ion sources within the Arches cluster \citep{Dogiel2009, Tatischeff2012}, which would increase the local CR ionization rate and therefore the LECRp contribution to the X-ray emission.
While it is uncertain when the Arches cluster X-ray emission will reach the constant background level, it is likely that Sgr~B2 has reached or will soon reach the background X-ray emission level. 
Therefore, the remaining emission from Sgr~B2 will be a unique powerful tool to probe the CR population in the GC in the X-ray energy bands.
The LECRp scenario will stay valid as long as no further decrease is observed in the coming years.

\subsection{The Nature of G0.66$-$0.13}
G0.66$-$0.13 is an elliptical feature with a major radius of $\sim5$~pc and a minor radius of $\sim3$~pc. 
The illuminating source photon index ($\Gamma=1.4\pm0.5$) is compatible with that of the central region of \sgrb\ ($\Gamma=2.2\pm0.4$). 
The intrinsic column density of $N_{H}(i)=3.0^{+3.8}_{-1.9} \times 10^{23}\rm~cm^{-2}$ is lower than the column density of the central part of \sgrb. 
The primary source luminosity based on its 2012 peak Fe K$\alpha$ emission is $L_{8}=(1.3\pm0.5) \times 10^{38} (R/\rm 100~pc)^2~(Z_{Fe})^{-1}\rm~erg~s^{-1}$, or $L_{3-79\rm~keV}=(4.3\pm1.6) \times 10^{38} (R/\rm~100 pc)^2~(Z_{Fe})^{-1}\rm~erg~s^{-1}$.
It is consistent with the primary source luminosity required by the central region of \sgrb.
The Fe K$\alpha$ line intensity of G0.66$-$0.13 exhibits fast decay at 6.4~keV. 
The linear fitting shows that the G0.66$-$0.13 Fe K$\alpha$ emission demonstrates an increasing trend prior to 2013 (Figure 5).
However, the increase could be much sharper than  what is shown in Figure 5.
We do not have good enough statistics to distinguish between a linear increasing from 2000 to 2012 and a sharp peak in 2012 on top of a baseline Fe K$\alpha$ emission.
In 2012, the peak Fe K$\alpha$ emission from G0.66$-$0.13 was brighter than the Sgr~B2 core Sgr~B2(M).
The non-detection of its Fe K$\alpha$ emission in 2013 suggests a fast decrease in Fe K$\alpha$ line flux.
The fast flux decrease with a life time of $t\sim 1$~year roughly matches the light crossing time of the peak within G0.66$-$0.13 ($r/c\sim2$~years).    

Both Sgr~B2 and G.66$-$0.13 are about 100~pc away from Sgr~A$^{\star}$, while the G0.66$-$0.13 X-ray flux reached the peak 12-18 years after Sgr~B2.
If G.66$-$0.13 was illuminated by the same Sgr~A$^{\star}$ outburst that illuminated Sgr~B2, it should be 16-23~pc behind the Sgr~B2 center in the line-of sight assuming Sgr~B2 is 130~pc in front of the projection plane \citep{Reid2009}. 
Therefore, G0.66$-$0.13 could be a molecular clump, with a higher local density than its surrounding environment, located in the \sgrb\ envelope which extends to about 22.5~pc.

\section{Conclusions}
We studied the X-ray emission of the central 90\asec\ of \sgrb\ and a newly discovered cloud feature G0.66$-$0.13 with \nustar. 
We present their broadband spectra, morphology and time variability and discuss two possible origins of the observed X-ray emission in 2013: X-ray reflection nebula and cosmic ray bombardment.  

$\bullet$ {\it Central 90\asec\ of Sgr~B2:} The substructures of \sgrb\ are resolved at sub-arcminute scales at hard X-ray energy ($>10$~keV). The hard X-ray continuum emission in 10--40~keV reveals two compact star-forming cores Sgr~B2(M) and Sgr~B2(N) surrounded by diffuse emission with the western side brighter than the eastern side. The central 90\asec\ region is marginally optically thin on average.

$\bullet$ {\it Main Compact Cores Sgr~B2(M) and Sgr~B2(N):} Compact cores Sgr~B2(M) and Sgr~B2(N) are resolved above 10 keV for the first time. Both the 6.4~keV line emission and the 10--40~keV continuum emission peak at the location of the main compact core Sgr~B2(M). While Sgr~B2(N) is significant in 10--40~keV continuum emission, it is not detected at 6.4~keV in 2013 perhaps due to higher local absorption. The Sgr~B2(M) spectrum requires a high optical depth of $\tau_{T}=1.3^{+2.5}_{-1.0}$, consistent with the {\it Herschel} measurement, suggesting it is not optically thin. 
 
$\bullet$ {\it XRN vs. LECR:} After a decade of decreasing, the 2012--2013 Fe K$\alpha$ emission reached 20\% of that in 2001, but remained at the same level during 2012--2013. The lack of data between 2005 and 2011 does not allow us to determine whether the 2013 Fe K$\alpha$ emission kept decreasing (which could be explained by the XRN or the LECRe model) or had reached the constant background level (for which the LECRp model is favored).
We first excluded the LECRe scenario based on unphysical best-fit parameters, no matter whether the Fe K$\alpha$ emission is decreasing or not. The significant 6.4~keV line of Sgr~B2 in 2013 requires the Fe abundance to be at least 3.4 solar, significantly exceeding the current measurements of the GC metallicity. The best-fit minimum electron energy is far below 100~keV. With such low energies, the non-thermal electrons are not able to penetrate \sgrb\ even if there is a local CR particle acceleration site. Therefore, we conclude that the LECRe scenario cannot be a major contributor to the remaining level of the X-ray emission from \sgrb\ in 2013. 

The 2013 Sgr B2 X-ray emission can be best explained by the XRN scenario if the X-ray emission is still decreasing.
We examine the XRN scenario with an ad hoc XRN model and a self-consistent XRN model. Due to lack of geometrical information in this ad hoc model, interpretation of the intrinsic column density of $N_{\rm H}(i)=(5.0\pm1.2)\times10^{23}\rm~cm^{-2}$ calls for caution. With the self-consistent XRN model {\tt MYTorus}, we derive the intrinsic equatorial column density to be $N_{H}(i)=(1.0\pm0.4) \times 10^{24}\rm~cm^{-2}$, twice that derived from the ad hoc model. This corresponds to $\tau_{T}=0.67\pm0.27$, confirming that the central 90\asec\ region is marginally optically thin on average. In the XRN scenario, we develope an analytical method to calculate the primary source luminosity via Compton scattering with the 10--40~keV scattered continuum (Equation 5). With this approach the primary source luminosity in any X-ray energy band can be derived with the observed continuum emission. The resulting primary source luminosity is $L_{3-79\rm~keV}(scat)=(5.6^{+5.2}_{-2.9})\times10^{38}(d/100\rm~pc)^2~erg~s^{-1}$. This is remarkably consistent with the primary source luminosity calculated via the photoelectric absorption process based on the updated source spectral shape, resulting in $L_{3-79\rm~keV}(abs)=(5.0\pm2.3)\times10^{38}(d/100\rm~pc)^2~erg~s^{-1}$. We find that $L_{8}(abs)/L_{8}(scat)$ is of order of unity, confirming the self-consistency of the XRN model. 

In case the Fe K$\alpha$ emission has reached the background level in 2013, the reflected X-rays could have completely faded and the LECRp process could be a major contributor. 
The required total CR ion power is $dW/dt=(0.6-3.2)\times10^{39}\rm~erg~s^{-1}$, about 10\% of the mechanical power supplied by supernovae in the inner $\sim200$~pc of the Galaxy. The CR ionization rate is found to be $\zeta_{H} \sim (6-10) \times 10^{-15}\rm~H^{-1}~s^{-1}$, consistent with the CR ionization rate in the GC environment. If the Sgr~B2 X-ray emission has indeed reached the background level, it would be a powerful tool to constrain the CR ion population in the GC.  

$\bullet$ {\it The new cloud feature G0.66$-$0.13:} The Fe K$\alpha$ line flux of G0.66$-$0.13 reached its peak in 2012 and quickly diminished within $\sim1$~year. The fast variability is best explained by the XRN scenario. The required primary source luminosity $L_{3-79\rm~keV}=(4.3\pm1.6) \times 10^{38} (d/\rm100pc)^2 Z_{Fe}^{-1}\rm~erg~s^{-1}$ is consistent with that derived from central Sgr~B2. Assuming G.66$-$0.13 is illuminated by the same Sgr~A$^{\star}$ outburst that illuminates Sgr~B2, it should be 16-23~pc behind the Sgr~B2 center along the line-of sight, which is still within the Sgr~B2 region. G0.66$-$0.13 could be a molecular clump with a local column density higher than surroundings located in the \sgrb\ envelope.

$\bullet$ {\it Other Galactic center molecular clouds and Sgr~A$^\star$ outburst history:} In the \nustar\ Galactic plane survey, other Galactic center giant molecular clouds were also detected well above 10~keV, including the ``Bridge" and G0.13$-$0.13. The same methodology to derive the primary source luminosity is applied to each cloud and reconstruct the Sgr~A$^\star$ outburst history (Mori et~al. submitted).

\acknowledgements
This work was supported under NASA Contract No. NNG08FD60C, and made use of data from the \nustar\ mission, a project led by the California Institute of Technology, managed by the Jet Propulsion Laboratory, and funded by NASA. We thank the \nustar\ Operations, Software and Calibration teams for support with the execution and analysis of these observations. This research has made use of the \nustar\ Data Analysis Software (NuSTARDAS) jointly developed by the ASI Science Data Center (ASDC, Italy) and the California Institute of Technology (USA). This research has also made use of data obtained with \xmm, an ESA science mission with instruments and contribution directly funded by ESA Member States and NASA. SZ is supported by NASA Headquarters under the NASA Earth and Space Science Fellowship Program - Grant ``NNX13AM31''. FEB acknowledges support from CONICYT-Chile and the Ministry of Economy, Development, and Tourism's Millennium Science Initiative. GP acknowledges support via an EU Marie Curie Intra European fellowship under contract no. FP-PEOPLE-2012-IEF-331095 and Bundesministerium f\"{u}r Wirtschaft und Technologie/Deutsches Zentrum f\"{u}r Luft-und Raumfahrt (BMWI/DLR, FKZ 50 OR 1408) and the Max Planck Society. MC, AG, RT and SS acknowledge support by CNES.

\appendix

\section{Applicability of {\tt MYTorus} model to Sgr~B2 Spectroscopy}

The {\tt MYTorus} model was originally developed to study X-ray spectra of Compton-thick AGNs.
The assumed geometry is a torus with a uniform distribution of neutral material reflecting incoming X-rays from an illuminating source at the center.
Three model components are offered: the transmitted continuum (MYTZ), the scattered continuum (MYTS) and Fe fluorescence lines (MYTL). 
The model allows a range of values for three key model parameters: the illuminating source photon index $\Gamma=1.4-2.6$, the equatorial hydrogen column density (corresponding to the minor diameter of the torus) $N_{H}=10^{22}-10^{25}\rm~cm^{-2}$ and the inclination angle $\theta_{obs}=0^{\circ}-90^{\circ}$. 
Considering a spherical molecular cloud as part of a virtual torus, the {\tt MYTorus} model can be applied to molecular cloud spectra with some limitations.
Figure A6 shows the geometry of a molecular cloud and a virtual torus. 
In the {\tt MYTorus} model, the observation angle $\theta_{obs}$ is defined as the angle between the light-of-sight (LOS) and the symmetry axis of the torus.
In contrast, most GC molecular cloud publications use the scattering angle $\theta$. 
A face-on case in the {\tt MYTorus} model ($\theta_{obs}=0^{\circ}$) corresponds to a cloud in the same projected plane as the illuminating source, where the scattering angle is $\theta=90^{\circ}$.
For three key assumptions of the {\tt MYTorus} model, we discuss the valid parameter space where the model is applicable to the molecular cloud X-ray reflection spectra in the following.

\begin{figure*} 
\centeringÊ 
\label{fig:mytorus}
\begin{tabular}{cc}
\includegraphics[angle=0, width=0.5\textwidth]{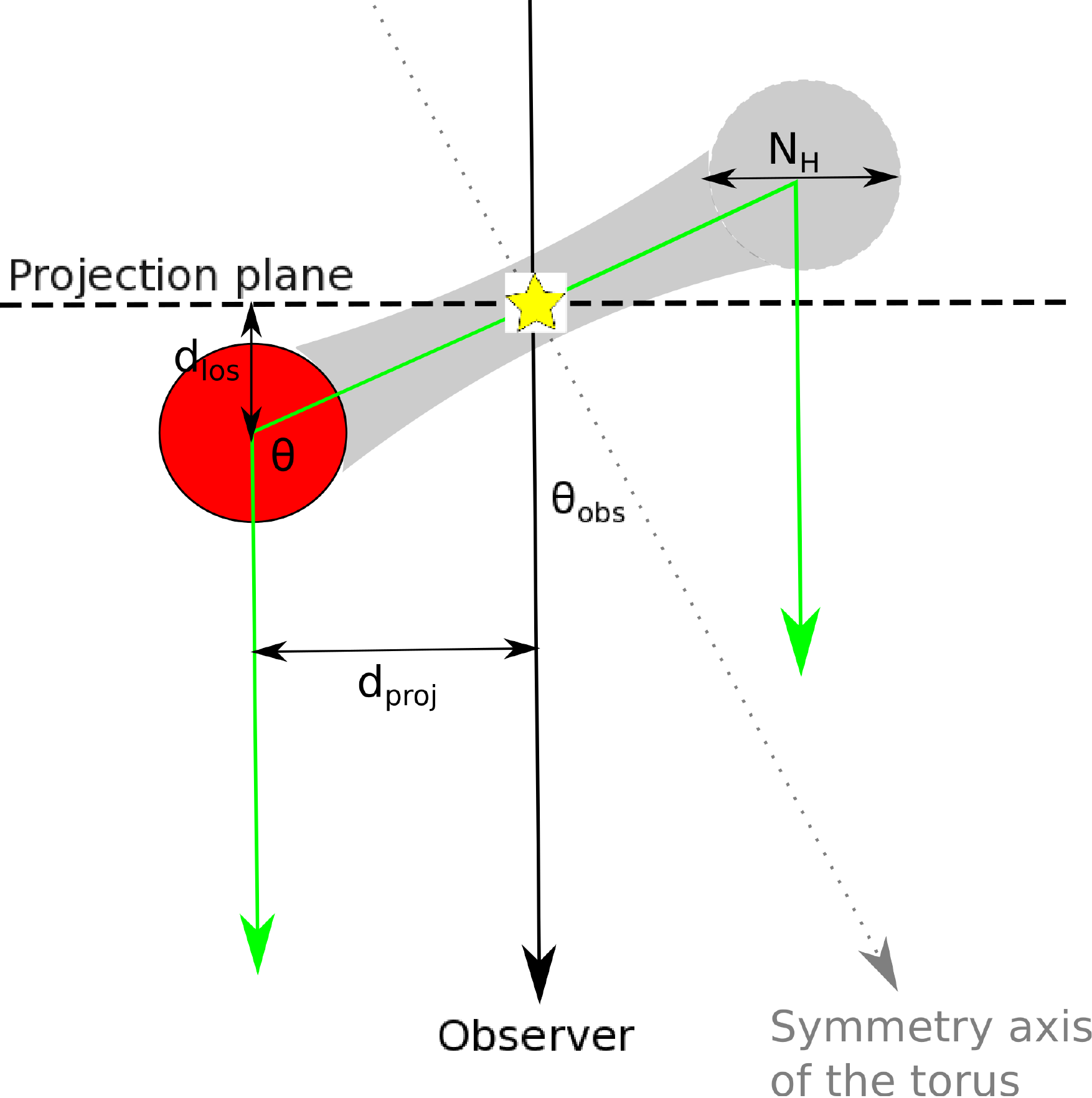} 
\end{tabular}
\caption{A spherical cloud (red circle) in a virtual torus (grey area) shows how the {\tt MyTorus} model geometry is related to the GCMC geometry. The illuminating source (Sgr~A$^\star$ for GCMC) is shown by the yellow star. 
In the {\tt MYTorus} model, the incident angle $\theta_{obs}$ is defined as the angle between the symmetry axis of the torus (dotted grey line) and the LOS (black vertical line). 
The equatorial column density $N_{H}$ is defined over the minor diameter of the torus.
When applying to the GCMC case, the quasi-spherical cloud can be considered as part of the torus.
Incoming X-ray photons from the illuminating source are scattered by the cloud at an angle $\theta$ to the direction of the observer.
The green lines show the travel paths of the X-ray photons.
$d_{proj}$ is the projected distance between the cloud and the illuminating source as seen by the observer, and $d_{los}$ is the LOS distance between the cloud and the projected plane.
A face-on view of the torus ($\theta_{obs}=0^{\circ}$) corresponds to the case where the cloud locates in the projected plane noted by a dashed black horizontal line ($\theta=90^{\circ}$, $d_{los}=0$).
The figure is from Appendix B1 in Mori et~al. (submitted).
}
\end{figure*}

\subsection{Toroidal Reflector}

In the face-on case ($\theta_{obs}=0^{\circ}$), {\tt MYTorus} gives an accurate solution for a fully illuminated quasi-spherical molecular cloud, as different azimuthal parts of the torus all scatter at the same angle $\theta=90^{\circ}$.
To derive the incoming X-ray flux that illuminates the cloud (red circle in Figure A6, which is part of the grey virtual torus), we simply need to rescale it by the solid angle ratio of the torus (fixed to $\Omega/2\pi=0.5$) and the cloud.
For Sgr~B2, its solid angle respect to Sgr~A$^{\star}$ is $\sim 2.89\times10^{-4}$. 
When the inclination angle deviates from $\theta_{obs}=0^{\circ}$, i.e. the face-on case, different azimuthal parts of the torus scatter incoming X-rays at different angles. 
The reflected spectrum thus shows variation and becomes inaccurate for a quasi-spherical molecular cloud.
However, the scattered component MYTS does not vary strongly with $\theta_{obs}$ as long as $\theta_{obs} \lesssim 60^{\circ}$ and $N_{H} \lesssim 10^{24}\rm~cm^{-2}$ (See Figure 17 in Mori et~al. 2015). 
When $\theta_{obs} > 60^{\circ}$, the spectral variation starts to become significant, for the following two reasons.
Firstly, as the inclination angle increases, the X-ray photons back-scattered by the  side of the torus hit the closer side of the torus before reaching the observer, thus are subject to further absorption.
Secondly, multi-scattering can become important and cause angular-dependent X-ray flux, although it is negligible at $N_{H} \lesssim 10^{24}\rm~cm^{-2}$.  

Next we determine the systematic error of measuring $N_{H}$, $\Gamma$ and model normalization based on the {\tt MYTorus} model.
We simulated MYTS spectra for $\theta=60^{\circ}$ and $N_{H}=10^{23}-10^{24}\rm~cm^{-2}$ and fit with the MYTS model with $\theta_{obs}$ set to $0^{\circ}$.
The deviations of the best-fit value of these model parameters from their input value are adopted as the systematic errors.
At $N_{H}=10^{23}\rm~cm^{-2}$, the deviation of $N_{H}$, $\Gamma$ and model normalization from their input are 10\%, 1\% and 7\%, while at $N_{H}=10^{24}\rm~cm^{-2}$, the deviations increase to 25\%, 3\% and 10\%.
Therefore we conclude that the reflected X-ray spectrum model {\tt MYTorus} is not sensitive to the reflector geometry in the \nustar\ energy band as long as $\theta_{obs} \lesssim 60^{\circ}$ and $N_{H} \lesssim 10^{24}\rm~cm^{-2}$.
Similar to the face-on case, the incident X-ray flux can be derived by rescaling it with the solid angle ratio of the torus $\Omega/4\pi=0.5$ and central 90\asec\ of Sgr~B2 cloud $\Omega/4\pi=2.89\times10^{-4} \times (100\rm~pc/R)^{2}$ with systematic error $<10\%$.
However, the conversion from the incident angle in the {\tt MYTorus} model to the position of the cloud cannot be well established.
We note that the $N_{H}$ measured by the {\tt MYTorus} model is an averaged value over the torus: it does not into account the specific geometry of the studied cloud and its possible partial illumination (see e.g. \citealp{Odaka2011}). The systematic errors are estimated solely based on the {\tt MYTorus} model and might therefore be underestimated. 
The more sophisticated model under development will better constrain the $N_{H}$ (Walls et al. in prep).

\subsection{Fe Abundance Fixed to Solar}

The Fe abundance is important in determining the incoming X-ray flux solely from measurements of Fe fluorescent lines,
while the 3--79~keV broadband spectrum of \nustar\ also provides the energy range where Compton scattering dominates ($>10\rm~keV$) over Fe fluorescence.
The incoming X-ray flux is determined consistently from both the Compton scattering process with the MYTS component and the Fe fluorescence process via the MYTL component.
The best-fit model parameters ($N_{H}$, $\Gamma$ and normalization) of MYTS do not vary significantly when the MYTS model is fit to the cloud X-ray spectra with or without the 6--10~keV~energy range which covers the 6.4~keV Fe K$\alpha$ line, the 7.06~keV Fe K$\beta$ line and the 7.1~keV Fe K edge.
Data tables for different Fe abundances in the range of $Z_{Fe}=0.5-3.0$ will be implemented to the modified XRN model for molecular clouds.
 
\subsection{Uniform Density}

The incoming X-ray flux from the illuminating source will not be significantly affected by non-uniform density profiles as long as $N_{H} \le 10^{24}\rm~cm^{-2}$, where multi-scattering is negligible.
If the density is not uniform, large clouds can include very dense clumps where multi-scattering effects can be significant.
For Sgr~B2, the largest optical depth is $\tau_{T} \sim 1$ in the direction of the densest cores Sgr~B2(M), in which case the multi-scattering effects are negligible. 
The current XRN model does not address the complicated density profile of Sgr~B2, which contains compact cores, clumps and an overall decreasing density profile. 
A more sophisticated XRN model with a reasonable density profile of Sgr~B2 implemented is under development (Walls et~al. in prep) and will reduce the uncertainties caused by non-uniformity of Sgr~B2.

\end{document}